\documentclass[10pt]{iopart}
\pdfoutput=1

\usepackage{iopams}
\usepackage{setstack}
\usepackage{graphicx}
\usepackage{float}
\usepackage{xcolor}

\usepackage{hyperref}
\hypersetup{bookmarks=True}

\begin{document}

\newcommand{\comjg}[1]{\textcolor{blue}{\textbf{[#1]}}}
\newcommand{\comeb}[1]{\textcolor{red}{\textbf{[#1]}}}

\newcommand{\eb}[1]{{\color{red} #1}}
\newcommand{\jg}[1]{{\color{black} #1}}

\newcommand{\p}{\partial}
\newcommand{\ri}{\mathrm{i}}
\newcommand{\re}{\mathrm{e}}
\newcommand{\bsf}[1]{\textsf{\textbf{#1}}}
\newcommand{\be}{\begin{equation}}
\newcommand{\ee}{\end{equation}}
\newcommand{\bea}{\begin{eqnarray}}
\newcommand{\eea}{\end{eqnarray}}
\newcommand{\ve}{\varepsilon}
\newcommand{\mC}{\mathcal{C}}
\newcommand{\mm}{\mathbf{m}}
\newcommand{\llbracket}{[}
\newcommand{\rrbracket}{]}

\newcommand{\ddr}{\ensuremath{\mathrm{d}}}     
\newcommand{\deriv}[2]{\ensuremath{\frac{\mathrm{d}#1}{\mathrm{d}#2}}}
\newcommand{\pderiv}[2]{\ensuremath{\frac{\partial#1}{\partial#2}}}
\newcommand{\average}[2]{\ensuremath{\left\langle {#1} \right\rangle_{#2}}}
\newcommand{\config}[0]{\ensuremath{\mathcal{C}}}

\newcommand{\vecbld}[1]{\ensuremath{\mathbf{#1}}}
\newcommand{\ps}[2]{\ensuremath{\langle #1 | #2 \rangle}}
\renewcommand{\vec}[1]{\ensuremath{|#1\rangle}}
\newcommand{\dvec}[1]{\ensuremath{\langle #1 |}}
\newcommand{\order}[2]{\ensuremath{\mathcal{O}({#2}^{#1})}}

\newcommand{\eqref}[1]{(\ref{#1})}

\title[]{Non-additive large deviations function for the particle densities of driven systems in contact}
\author{Jules Guioth$^1$ and Eric Bertin$^2$}
\address{$^1$
DAMTP, Centre for Mathematical Sciences, University of Cambridge, Wilberforce Road, Cambridge CB3 0WA, UK}
\address{$^2$
Univ.~Grenoble Alpes and CNRS, LIPHY, F-38000 Grenoble, France}

\ead{jules.guioth@damtp.cam.ac.uk, eric.bertin@univ-grenoble-alpes.fr}

\begin{abstract}
We investigate the non-equilibrium large deviations function of the particle densities in two steady-state driven systems exchanging particles at a vanishing rate. We first derive through a systematic multi-scale analysis the coarse-grained master equation satisfied by the distribution of the numbers of particles in each system. Assuming that this distribution takes for large systems a large deviations form, we obtain the equation (similar to a Hamilton-Jacobi equation) satisfied by the large deviations function of the densities. Depending on the systems considered, this equation may satisfy or not the macroscopic detailed balance property, i.e., a time-reversibility property at large deviations level.
In the absence of macroscopic detailed balance, the large deviations function can be determined as an expansion close to a solution satisfying macroscopic detailed balance. In this case, the large deviations function is generically non-additive, i.e., it cannot be split as two separate contributions from each system.
In addition, the large deviations function can be interpreted as a non-equilibrium free energy, as it satisfies a generalization of the second law of thermodynamics, in the spirit of the Hatano-Sasa relation.
Some of the results are illustrated on an exactly solvable driven lattice gas model.
\end{abstract}


\section{Introduction}

A key issue of thermodynamics is to be able to determine the steady-state values of particle density or energy density for instance, in two systems brought in contact so that particles or energy can be exchanged between them. At equilibrium, such densities of globally conserved quantities are determined by the equalization in both systems of the conjugated intensive thermodynamic parameter, namely chemical potential or temperature.
Out of equilibrium, the situation is generically more complex, but it is natural to try to generalize the equilibrium formalism and to try to define non-equilibrium intensive parameters like temperature, chemical potential and pressure \cite{Oono1998,sasa2006steady}. In particular, the non-equilibrium extension of the notion of temperature in steady-state systems has raised a lot of attention \cite{Jou03,Cugliandolo11,Levine07,Bertin04,Martens09}, but due to the lack of energy conservation in driven dissipative system, a thermodynamic consistent definition of a non-equilibrium temperature remains elusive in most cases \cite{Dickman2014Inconsistencies}.
Yet, particle number or volume may still be conserved out of equilibrium, and it is tempting to try to define chemical potential or pressure in this situation.
Pressure may actually be defined from a purely mechanical standpoint in terms of normal force per unit surface on the confining walls, as recently done for gases of active particles \cite{solon2015nat,solon2015prl,winkler2015virial,Joyeux2016,speck2016ideal,fily2018mechanical}.
This definition of pressure, though simple and unambiguous, generically leads to the lack of an equation of state \cite{solon2015nat} (except if specific symmetries are present \cite{solon2015prl}), meaning that the pressure no longer depends only on the bulk density of the gas, but also on the detailed properties of the confining walls.

In contrast, the chemical potential cannot be defined from purely mechanical considerations, and one needs to come back, in analogy to the equilibrium definition, to a definition based on a statistical characterization of density fluctuations in two subsystems exchanging particles. Such an approach actually relies on a additivity property of the large deviations function describing density fluctuations in subsystems \cite{bertin2006def,bertin2007intensive,pradhan2010nonequilibrium,pradhan2011approximate,chatterjee2015zeroth}.
Recent works have more explicitly focused on the properties of the non-equilibrium chemical potential, which turns out to generically depend on the contact dynamics and to lack an equation of state \cite{guioth2018large,guioth2019lack,GuiothTBP19}. The validity of the additivity condition has been traced back to both a macroscopic detailed balance condition at contact and a factorization of the contact dynamics \cite{guioth2018large,GuiothTBP19}.

When these conditions are not satisfied, the large deviations function is not additive and does not decompose as a sum of two contributions depending only on one of the two systems. It is thus not possible to define a non-equilibrium chemical potential. However, the large deviations function of the densities still contains the relevant statistical information to characterize the steady-state densities in the two systems, as well as their fluctuations. In this work, we present a detailed study of the large deviations function of the densities in two systems in \jg{weak} contact, focusing on the case where the large deviations function is not additive so that no chemical potential can be defined. We show how this large deviations function can be evaluated, and we provide it with a thermodynamic interpretation in terms of a generalized second law of thermodynamics, in the spirit of the Hatano-Sasa relation \cite{hatano2001steady,bertini2015macroscopic}.

The paper is organized as follows.
Sec.~\ref{sec:contact} introduces the set-up of two driven systems in contact and derives the equation governing the evolution of the distribution of densities in both systems.
Sec.~\ref{sec:largedev} performs a large deviations analysis of the steady-state distribution of densities, and discusses the notion of macroscopic detailed balance. A perturbative evaluation of the large deviations function in the absence of macroscopic detailed balance is presented, and the consequences regarding the additivity property are discussed.
Sec.~\ref{sec:potential} deals with the thermodynamic interpretation of the large deviations function in terms of a generalized second law of thermodynamics. Finally, Sec.~\ref{sec:MTM} presents an application of this large deviations formalism to an exactly solvable driven lattice model, in which particles (or possibly continuous amounts of `mass') are transported on a one-dimensional lattice.


\section{Dynamics of two driven systems in contact}
\label{sec:contact}

\subsection{Contact dynamics}

\subsubsection{Stochastic driven lattice models}
\label{sec:def:dyn}

Throughout this paper, we consider stochastic lattice gases (or generalizations involving continuous masses instead of particles) \cite{liggett2012interacting,spitzer1970}, that are stochastic jump processes in which particles randomly jump from site to site on a lattice. This include for instance the Zero Range Process \cite{evans2005nonequilibrium,Levine2005} and its generalizations \cite{Evans2006,evans2004factorized,evans2006factorized,zia2004construction}, the Asymmetric Simple Exclusion Process (ASEP) \cite{spitzer1970,derrida2007non,derrida1998asep}, or the Katz-Lebowitz-Spohn (KLS) model \cite{katz1984nonequilibrium,zia2010twenty}.

We generically denote as $\Lambda \subset \mathbb{Z}^{\mathrm{d}}$ the $d$-dimensional lattice, $V=|\Lambda|$ the number of sites, $N$ the total number of particles. A microscopic configuration of the model is written as $\config = \{ n_{x} \}_{x \in \Lambda}$, where $n_{x}\in [0, n_{\mathrm{max}}]$ is the number of particles at site $x$. Note that the maximal number of particles $n_{\mathrm{max}}$ on a single site can be finite or infinite.
In most models, the local configuration $n_{x}$ is an integer, but it might also be a real variable $n_{x} \geq 0$, in which cases it may be called a `continuous mass' \cite{Evans2006,evans2004factorized,evans2006factorized,zia2004construction}. To allow for stationary particle flux along the drive, periodic boundary conditions are assumed at least in this direction.

The probability per unit time that a particle jumps from one site to another is given by the transition rates $T(\config^{\prime}|\config)$, which are expressed in terms of the configurations $\config$ and $\config^{\prime}$ respectively before and after the jump.
For thermodynamic consistency, one imposes the \emph{local detailed balance} \cite{katz1984nonequilibrium,maes2003origin,maes2003time,wynants2010structures} condition, which reads
\begin{equation}
  \label{eq:local_DB}
  \frac{T(\config^{\prime}|\config)}{T(\config|\config^{\prime})} = \exp \left[ -\beta \left(E(\config^{\prime})-E(\config) - W_{\mathrm{nc}}(\config,\config^{\prime}) \right) \right]
\end{equation}
where $E(\config)$ denotes the energy associated with the configuration $\config$ (composed of an interaction potential and an external potential) and $W_{\mathrm{nc}}(\config,\config^{\prime})$ is the non-conservative work generated by the drive.
This non-conservative work by definition depends on the drive, and we assume for the sake of simplicity a constant driving force $f$, leading to
\begin{equation}
    W_{\mathrm{nc}}(\config,\config^{\prime})= f \cdot j(\config,\config^{\prime}),
\end{equation}
where $j(\config,\config^{\prime})$ is the particle current associated with the transition $\config \to \config^{\prime}$.
This constraint, though, is not enough to fully specify the transition rates $T(\config^{\prime}|\config)$. We will see in the following several standard choices 
obeying local detailed balance, e.g., the Metropolis rule, the Kawasaki rule, the exponential rule, or the Sasa-Tasaki rule \cite{tasaki2004remark}.

\subsubsection{Contact dynamics between two systems}
\label{sec:def:contact}

Our aim in this work is to study driven systems in contact. To proceed further, we need to specify a generic framework for the contact dynamics.
On generic grounds, we consider two systems $A$ and $B$ characterized by the energies $E_{A}(\config_{A})$, $E_{B}(\config_{B})$ associated with their configurations $\config_{A}$ $\config_{B}$. Each system is subjected to a driving force, $f_{A}$ or $f_{B}$.
Using the notations introduced above, we note $\Lambda_{k}$ the lattice of system $k$, $V_{k}=|\Lambda_{k}|$ the corresponding number of sites, and $N_{k}=\mathcal{N}(\config_{k})$ the number of particles in system $k$, $k=A,B$.
We also introduce the geometric factors $\gamma_{A}=V_{A}/V$ and $\gamma_{B}=V_{B}/V$.

We focus here on the situation where the contact is orthogonal to the driving forces. Hence we assume that the microscopic transition rate $T_{c}$ at contact does not depend on the driving forces $f_{A}$, $f_{B}$.
The stochastic dynamics at contact is thus defined by a transition rate $T_{\mathrm{c}}(\config_{A}^{\prime},\config_{B}^{\prime}|\config_{A},\config_{B})$, which we assume to obey local detailed balance with respect to the equilibrium distributions of systems $A$ and $B$. The contact dynamics only involves particle exchange between the two systems, so that the total number of particles $N=\mathcal{N}(\config_{A})+\mathcal{N}(\config_{B})$ is conserved.

The master equation governing the evolution of the probability $P_{t}(\config_{A},\config_{B})$ describing the statistics of the two systems in contact thus reads
\begin{eqnarray}
  \label{eq:micro_master_equation_composed_system_AB}
  && \deriv{P_{t}}{t}(\config_{A},\config_{B})  = \\
  && \hspace{1em} \sum_{\config_{A}'\neq \config_{A}} T_{A}(\config_{A}|\config_{A}') P_{t}(\config_{A}',\config_{B}) - \lambda_{A}(\config_{A})P_{t}(\config_{A},\config_{B})  \nonumber \\
  && \hspace{1.5em} + \sum_{\config_{B}' \neq \config_{B}} T_{B}(\config_{B}|\config_{B}') P_{t}(\config_{A}, \config_{B}') - \lambda_{B}(\config_{B})P_{t}(\config_{A},\config_{B}) \nonumber \\
  && \hspace{1.5em} + \sum_{\config_{A}' \neq \config_{A},\, \config_{B}' \neq \config_{B}} T_{c}(\config_{A},\config_{B}|\config_{A}',\config_{B}') P_{t}(\config_{A}',\config_{B}') \nonumber \\
  && \hspace{12em} - \lambda_{c}(\config_{A},\config_{B}) P_{t}(\config_{A},\config_{B}) \; \nonumber
\end{eqnarray}
where $\lambda_{k}(\config)=\sum_{\config'\neq\config}T_{k}(\config'|\config)$ 
denotes the escape rate, $k$ being $A,\,B$ or $c$.

In the following, we assume that the transition rate at contact is small to allow for a time scale separation between bulk and contact dynamics. To make this time scale separation explicit, we now write the transition rate at contact as
$\epsilon T_{c}$, where $\epsilon$ is a small parameter.

\subsection{Time-scale separation in the weak contact limit: multi-scale analysis}
\label{sec:multi-scale_analysis}

In this section, we take advantage of the time scale separation to perform a multi-scale analysis of the master equation  (\ref{eq:micro_master_equation_composed_system_AB}) \cite{nayfeh2008perturbation,bender1999advanced,chen1996renormalization,chen1994renormalization}.
With this aim in mind, we first lighten notations and rewrite the master equation (\ref{eq:micro_master_equation_composed_system_AB}) using more formal vector notations:

\begin{equation}
  \label{eq:master_eq_config}
  \deriv{\vec{P_{t}}}{t} = \mathcal{W}_{b}\vec{P_{t}} + \epsilon \mathcal{W}_{c} \vec{P_{t}} 
\end{equation}
where $\mathcal{W}_{b}$ and $\mathcal{W}_{c}$ are respectively the evolution matrices associated with the bulk transition rates $T_{b}(\mC|\mC^{\prime}) = T_{A}(\mC_{A}|{\mC_{A}}^{\prime}) \delta_{\mC_{B}, \mC_{B}'} + T_{B}(\mC_{B}|{\mC_{B}}^{\prime}) \delta_{\mC_{A},\mC_{A}'}$ and the contact transition rates $T_{c}(\mC|\mC^{\prime})$. $\vec{P_{t}}$ is the vector whose coordinates are $P_{t}(\mC)=\ps{\mC}{P_{t}}$, $\dvec{\mC}$ being the row vector associated with the configuration $\mC$ (full of $0$ except at the configuration label $\mC$ for which it is equal to $1$).  

To get the solution of this master equation \eqref{eq:master_eq_config} in the weak contact limit $\epsilon \to 0$, we perform a perturbative expansion
\begin{equation}
  \label{eq:perturbative_exp_vec_Pt}
  \vec{P_{t}} = \vec{P_{t}^{(0)}} + \epsilon \vec{P_{t}^{(1)}} + \Or\left(\epsilon^{2}\right) \, .
\end{equation}
The master equation reads, at order $\Or\left(\epsilon^{0}\right)$ and $\Or\left(\epsilon^{1}\right)$
\begin{eqnarray}
  \label{eq:master_eq_coupled_eps_orders}
  \Or\left(\epsilon^{0}\right)\, : \qquad && \deriv{\vec{P_{t}^{(0)}}}{t} = \mathcal{W}_{b}\vec{P_{t}^{(0)}} \\
  \Or\left(\epsilon^{1}\right)\, : \qquad && \deriv{\vec{P_{t}^{(1)}}}{t} = \mathcal{W}_{b}\vec{P_{t}^{(1)}} + \mathcal{W}_{c}\vec{P_{t}^{(0)}} \, . \nonumber
\end{eqnarray}
These two last equations admit the formal solutions
\begin{eqnarray}
  \label{eq:sol_master_eq_eps_orders}
  \vec{P_{t}^{(0)}} && = e^{t\mathcal{W}_{b}}\vec{P_{0}} \\
  \vec{P_{t}^{(1)}} && = \int_{0}^{t}\!\!\! \mathrm{d}s \, e^{(t-s)\mathcal{W}_{b}}\mathcal{W}_{c}e^{s\mathcal{W}_{b}}\vec{P_{0}} \;, \nonumber
\end{eqnarray}
with initial conditions $\vec{P_{t=0}}=\vec{P_{0}}$, where $\vec{P_{0}} \sim \Or\left(\epsilon^{0}\right)$. The initial condition is arbitrary and can generically be decomposed as
\begin{equation}
  \label{eq:initial_condition}
  \vec{P_{0}} = \sum_{\rho_{A}}P_{0}(\rho_{A})\vec{SS,\rho_{A}}_{b} + \sum_{\rho_{A}}\vec{T_{0},\rho_{A}}_{b}
\end{equation}
where $\vec{SS,\rho_{A}}_{b}$ is a short notation for a stationary solution of the bulk dynamics with densities $\rho_{A}$ and $\rho_{B}$ chosen initially: $\ps{\mC}{SS,\rho_{A}}_{b} = P_{A}(\mC_{A}|\rho_{A})P_{B}(\mC_{B}|\rho_{B})$ for any $\mC=(\mC_{A},\mC_{B})$ such that $\mathcal{N}(\mC_{A})=\rho_{A}V_{A}$ and $\mathcal{N}(\mC_{B})=\rho_{B}V_{B}=N-\rho_{A}V_{A}$. To lighten notations, we omit the fixed total density $\bar{\rho}$. Technically, $\vec{SS,\rho_{A}}_{b}$ is a right eigenvector of $\mathcal{W}_{b}$ associated with the eigenvalue $0$. As for $\vec{T_{0},\rho_{A}}_{b}$, it refers to a transient part that vanishes at large time ($\lim_{t\to\infty}e^{t\mathcal{W}_{b}}\vec{T_{0},\rho_{A}}_{b} = 0$).
Finally, $P_{0}(\rho_{A})$ weights the probability to start with densities $\rho_{A}, \, \rho_{B}$ in the steady states of $A$ and $B$.

Inserting $\vec{P_{0}}$ in equations~\eqref{eq:sol_master_eq_eps_orders}, one obtains 
\begin{eqnarray}
  \label{eq:sol_master_eq_eps_orders_detailed}
  \vec{P_{t}^{(0)}} && = \sum_{\rho_{A}} P_{0}(\rho_{A})\vec{SS,\rho_{A}}_{b} + \sum_{\rho_{A}} e^{t\mathcal{W}_{b}}\vec{T_{0},\rho_{A}}_{b} \\
  \vec{P_{t}^{(1)}} && = t \sum_{\rho_{A},\rho_{A}^{\prime}} P_{0}(\rho_{A})\, {}_{b}{\dvec{-,\rho_{A}^{\prime}}} \mathcal{W}_{c}\, \vec{SS,\rho_{A}}_{b}\, \vec{SS,\rho_{A}^{\prime}}_{b} \nonumber \\
                    && \qquad + \sum_{\rho_{A}} P_{0}(\rho_{A}) \int_{0}^{t}\!\! \mathrm{d}s \, e^{(t-s)\mathcal{W}_{b}}\vec{R_{SS},\rho_{A}}_{b} \nonumber \\
                    && \qquad + \sum_{\rho_{A}} \int_{0}^{t}\!\! \mathrm{d}s \, e^{(t-s)\mathcal{W}_{b}} \mathcal{W}_{c} e^{s\mathcal{W}_{b}} \vec{T_{0},\rho_{A}}_{b} \nonumber
\end{eqnarray}
where we have used $\mathcal{W}_{c}\vec{SS,\rho_{A}}_{b} = \sum_{\rho_{A}^{\prime}}\, {}_{b}\dvec{-,\rho_{A}}\mathcal{W}_{c}\vec{SS,\rho_{A}}_{b} \, \vec{SS,\rho_{A}^{\prime}}_{b} + \vec{R_{SS},\rho_{A}}_{b}$, $\vec{R_{SS},\rho_{A}}_{b}$ being the transient component (with respect to the bulk dynamics) and ${}_{b}\dvec{-,\rho_{A}}$ being the left eigenvector of $\mathcal{W}_{b}$ for fixed densities $\rho_{A},\,\rho_{B}$ with eigenvalue 0. One has ${}_{b}\ps{-,\rho_{A}}{\mC} = 1$ if $\mathcal{N}_{A}(\mC) = V_{A}\rho_{A}$ and ${}_{b}\ps{-,\rho_{A}}{\mC} = 0$ otherwise.

The second and the third term of $\vec{P_{t}^{(1)}}$ in equation ~\eqref{eq:sol_master_eq_eps_orders_detailed} converge when $t\to\infty$ but the first, proportional to $t$, is clearly a \emph{secular} term which breaks the validity of the perturbation expansion as soon as $\epsilon t \sim \Or(1)$: for $t \sim \Or\left(\epsilon^{-1}\right)$, $\epsilon \vec{P_{t}^{(1)}}$ becomes of the same order as $\vec{P_{t}^{(0)}}$ and the expansion is no more uniform \cite{nayfeh2008perturbation,bender1999advanced}. Such discrepancy is the consequence of the fact that the perturbation series is slowly convergent and that all terms are needed to obtain a bounded result for any time $t$. In order to regularise the perturbation series for large time, one can use the fact that the initial condition cannot be observed when $t$ is large \cite{chen1994renormalization,chen1996renormalization,oono2012nonlinear}, and take advantage of this fact to renormalize the series.

In order to perform this multi-scale analysis, one introduces an arbitrary time $\tilde{t}$ which is interpreted as the new initial time, by writing $t=(t-\tilde{t}) + \tilde{t}$. Absorbing the term proportional to $\epsilon \tilde{t}$ of the secular term in the coefficients $P_{0}(\rho_{A})$, which leads to a renormalized term $P_{\tilde{t}}(\rho_{A})$, allows us to write the solution $\vec{P_{t}}$ (equation~\eqref{eq:perturbative_exp_vec_Pt}) as
\begin{eqnarray}
  \label{eq:sol_master_eq_renormalized_expansion}
  \vec{P_{t}} && = \sum_{\rho_{A}} P_{\tilde{t}}(\rho_{A})\vec{SS,\rho_{A}}_{b} + \sum_{\rho_{A}} e^{t\mathcal{W}_{b}}\vec{T_{0},\rho_{A}}_{b}  \\
              && \quad + \epsilon(t-\tilde{t}) \sum_{\rho_{A},\rho_{A}^{\prime}} P_{\tilde{t}}(\rho_{A})\, {}_{b}{\dvec{-,\rho_{A}^{\prime}}} \mathcal{W}_{c}\, \vec{SS,\rho_{A}}_{b}\, \vec{SS,\rho_{A}^{\prime}}_{b} \nonumber \\
             && \qquad + \epsilon \mathcal{R}_{t} + \Or\left(\epsilon^{2}\right)  \nonumber
\end{eqnarray}
where $\mathcal{R}_{t}$ refers to non-secular terms at order $\Or\left(\epsilon^{1}\right)$ in equation~\eqref{eq:perturbative_exp_vec_Pt}. The secular term $\propto \epsilon(t-\tilde{t})$ can now be deleted by a suitable choice of $P_{\tilde{t}}$. Indeed, decomposing $P_{\tilde{t}} = P_{t} + (t-\tilde{t})\partial P_{\epsilon t}/\partial t + o\left((t-\tilde{t})\right)$ in equation~\eqref{eq:sol_master_eq_renormalized_expansion}, one observes that -- after a projection on ${}_{b}\dvec{-,\rho_{A}}$~-- the secular term of order $\epsilon(t-\tilde{t})$ can be removed if $P_{t}$ satisfies
\begin{equation}
  \label{eq:slow_renormalized_eq_density}
  \deriv{P_{t}(\rho_{A})}{t} = \epsilon \sum_{\rho_{A}'} \, \pi(\rho_{A}|\rho_{A}') P_{t}(\rho_{A}') - \pi(\rho_{A}'|\rho_{A}) P_{t}(\rho_{A})
\end{equation}
where one has introduced,
\begin{equation}
  \label{eq:detail_transrate}
  \pi(\rho_{A}'|\rho_{A}) - \lambda(\rho_{A})\delta_{\rho_{A},\,\rho_{A}^{\prime}} = {}_{b}{\dvec{-,\rho_{A}^{\prime}}} \mathcal{W}_{c}\, \vec{SS,\rho_{A}}_{b} \, ,
\end{equation}
with $\rho_{A}^{\prime} = \rho_{A} + \Delta N_{A}/V_{A}$, $\pi(\rho_{A}'|\rho_{A}) $ being the transition rate associated with the transition from $\rho_{A}$ to $\rho_{A}^{\prime}$ ($\pi(\rho_{A},\rho_{A})=0$) and $\lambda(\rho_{A}) = \sum_{\rho_{A}'} \pi(\rho_{A}'|\rho_{A})$, the escape rate.

The final regularised solution eventually reads
\begin{eqnarray}
  \label{eq:final_Pt_regularized}
  \vec{P_{t}} && = \sum_{\rho_{A}} P_{\epsilon t}(\rho_{A}) \vec{SS,\rho_{A}}_{b} +  \sum_{\rho_{A}} e^{t\mathcal{W}_{b}}\vec{T_{0},\rho_{A}}_{b} \\
              && \qquad + \Or\left(\epsilon\right) \, . \nonumber
\end{eqnarray}
where the probability distribution $P_{\tau}(\rho_{A})$ obeys the coarse-grained master equation \eqref{eq:slow_renormalized_eq_density}:
\begin{equation}
  \label{eq:coarse_grained_master_eq_multiscale_analysis}
  \deriv{P_{\tau}}{\tau}(\rho_{A}) = \sum_{\rho_{A}'} \pi(\rho_{A}|\rho_{A}') P_{\tau}(\rho_{A}') - \pi(\rho_{A}'|\rho_{A}) P_{\tau}(\rho_{A}) \; ,
\end{equation}
with $\tau = \epsilon t$, the relevant slow time associated with the dynamics of the number of particles. 

For large time compared to the relaxation time of the bulk dynamics, the stationary solution $\lim_{t\to\infty}\vec{P_{t}} = \vec{P}$ reads, for all configurations $\mC$,
\begin{equation}
  \label{eq:stationary_solution_coupled_syst}
  P(\mC) = P(\rho_{A}|\bar{\rho}) P_{A}(\mC_{A}|\rho_{A})P_{B}(\mC_{B}|\rho_{B}) + \Or\left(\epsilon\right) \, ,
\end{equation}
with $P(\rho_{A}|\bar{\rho})$ the stationary solution\footnote{The dependence with respect to the total density has been reintegrated to not forget that $P$ describes the density $\rho_{A}$ in $A$ as well as in $B$, with $\rho_{B}=\gamma_{B}^{-1}(\bar{\rho}-\gamma_{A}\rho_{A})$.} of equation~\eqref{eq:coarse_grained_master_eq_multiscale_analysis}. 

In many cases of physical relevance, only one or at least a finite number $\Delta N_A$ of particles can be simultaneously exchanged at contact. In that case, the coarse-grained transition rate $\pi(\rho_{A}'|\rho_{A})$ takes for large system size the simple form
\begin{equation} \label{eq:pi:varphi}
\pi(\rho_{A}'|\rho_{A}) \underset{V_{A}\to\infty}{\to} \varphi(\rho_A,\Delta N_A)
\end{equation}
where $\rho_{A}'-\rho_{A}=\Delta N_A/V_A$.
A more detailed discussion of this point, in particular regarding the scaling assumptions underlying Eq.~\eqref{eq:pi:varphi}, can be found in \cite{GuiothTBP19}.
For later convenience, we repeat here the expression of the coarse-grained transition rates $\varphi(\rho_A,\Delta N_A)$, using explicit notations:
  \begin{equation}
    \label{eq:detailed_coarse-grained_transrate_contact}
    \!\!\!\!\!\!\!\!\!\!\!\!\!\!\!\!\!\!\!\!\!
     \varphi(\rho_A,\Delta N_A) = {\hat{\sum_{\mC_{A}', \mC_{B}'}}}^{\!\!\!(\Delta N_A)} \hat{\sum_{\mC_{A},\mC_{B}}}^{\!\!\!(0)}
    T_{c}(\mC_{A}', \mC_{B}' | \mC_{A}, \mC_{B} ) \, P_{A}(\mC_{A}|\rho_{A}) P_{B}(\mC_{B} | \rho_{B}) \, ,
  \end{equation}
  where $P_{k}(\mC_{k}|\rho_{k})$ is the stationary density of the isolated system $k$, and with the shorthand notation $\hat{\sum}_{\mC_{A},\mC_{B}}^{(\Delta N_A)}$ that stands for the constrained sum over $(\mC_{A}, \mC_{B})$ such that $\mathcal{N}(\mC_{A})=V_{A}\rho_{A} + \Delta N_{A}$ and $\mathcal{N}(\mC_{B})=V_{B}\rho_{B} - \Delta N_{A}$.

\section{Large deviations analysis of the density dynamics}
\label{sec:largedev}

\subsection{Hamilton Jacobi equation}

In order to study the large size limit, $V_A$, $V_B \to \infty$, it is natural to assume a large deviations form for the steady-state distribution $P(\rho_A)$,
\begin{equation}
  \label{eq:large_dev_ansatz}
  P(\rho_{A}|\bar{\rho}) \asymp e^{-V_A I(\rho_{A}|\bar{\rho})} \; ,
\end{equation}
where $\asymp$ refers to a logarithmic equivalence \cite{touchette2009large}.
Using this form in the coarse-grained master equation (\ref{eq:coarse_grained_master_eq_multiscale_analysis}) leads, in the limit $V_A$, $V_B \to \infty$, to the so-called Hamilton-Jacobi equation \cite{maes2007static}. In the stationary regime, the latter reads as
\begin{equation}
   \label{eq:hamilton-jacobi_eq}
   \sum_{\Delta N_{A}\neq 0} \varphi(\rho_{A},\Delta N_{A})\left[ e^{\Delta N_{A} I'(\rho_{A}|\bar{\rho})} - 1 \right] = 0
\end{equation}
A more detailed discussion on the derivation of this Hamilton-Jacobi equation starting from the coarse-grained master equation (\ref{eq:coarse_grained_master_eq_multiscale_analysis}) can be found in \cite{GuiothTBP19}.

\subsection{Macroscopic detailed balance}
\label{sec:macro_detailed_balance}

\subsubsection{Formal approach}

We now focus on the steady state of the Hamilton-Jacobi equation \eqref{eq:hamilton-jacobi_eq}.
This  steady state can be easily determined if each term under the following rearranged sum separately cancels for any value of $\rho_{A}$:
\begin{eqnarray} \label{eq:macroDB}
  && \sum_{\Delta N_{A}\neq 0} \varphi(\rho_{A},\Delta N_{A})\left[ e^{\Delta N_{A} I'(\rho_{A}|\bar{\rho})} - 1 \right] \\
  && \qquad = \sum_{\Delta N_{A}\neq 0}\underbrace{\left[\varphi(\rho_{A},\Delta N_{A}) e^{\Delta N_{A} I'(\rho_{A}|\bar{\rho})} - \varphi(\rho_{A},-\Delta N_{A})\right]}_{=0\; \mbox{if detailed balance}} = 0 \; . \nonumber
\end{eqnarray}
Formally, one gets in this way a generalised detailed balance condition, that we call \emph{macroscopic detailed balance} in the following.
Solving for $I'$, the detailed balance condition reads as
\begin{equation}
  \label{eq:macro_detailed_balance}
  I^{\prime}(\rho_{A}|\bar{\rho}) = \frac{1}{\Delta N_{A}} \ln \frac{\varphi(\rho_{A},-\Delta N_{A})}{\varphi(\rho_{A},\Delta N_{A})} \; .
\end{equation}

Quite importantly, note that for most lattice gas models~-- that deal with the dynamics of particles on lattice in continuous time~-- (and potentially also for more realistic systems), only one particle can be exchanged per unit time. Thus $\Delta N_{A} = \pm 1$ at most and one can easily check that the macroscopic detailed balance condition is always verified. However, for more general situations where $\max(\Delta N_{A}) \geq 2$ (or if $\Delta N_{A}$ is a continuous real quantity), this condition is not met in general. 
Before discussing how to deal with this more general case where macroscopic detailed balance is broken, we first discuss in subsections \ref{sec:sub:time-reversal} and \ref{sec:force_and_activity_transrate} the connection between the formal macroscopic detailed balance condition introduced in \eqref{eq:macroDB} and time-reversal symmetry.
This discussion also allows us to introduce some definitions and notations that will prove useful in the sequel.

\subsubsection{Time-reversal symmetry}\label{sec:sub:time-reversal}

Apparently, the macroscopic detailed balance condition~\eqref{eq:macroDB} is reminiscent of a time-reversal symmetry.
Indeed, the usual detailed balance is nothing but the equality, for all trajectories, between the probability to observe a certain trajectory and the probability to observe its time-reversed counterpart. In particular, the time-reversal symmetry for a two-time infinitesimal trajectory (on the time interval $[t,t+\ddr t]$) of the density $\rho_{A}(t)$ reads as
\begin{eqnarray}
  \label{eq:usual_detailed_balance}
  P(\rho_{A}', t+\ddr t ; \rho_{A}, t) && = P(\rho_{A}, t+\ddr t ; \rho_{A}', t) \\
  \pi(\rho_{A}'| \rho_{A})  P(\rho_{A}|\bar{\rho}) &&= \pi(\rho_{A}|\rho_{A}') P(\rho_{A}'|\bar{\rho}) \, . \nonumber
\end{eqnarray}
Thus, if $P(\rho_{A}|\bar{\rho}) \sim e^{-V_{A}{I}(\rho_{A}|\bar{\rho})}$, and recalling that $\rho_{A}'=\rho_{A} + \Delta N_{A}/V_{A}$, one gets at leading order in $V_{A}$ the macroscopic detailed balance \eqref{eq:macro_detailed_balance}.

We must stress here that the macroscopic detailed balance relation may hold without requiring that the microscopic, local, detailed balance holds as well. Indeed, the ratio between the probability to observe a transition $\mC=(\mC_{A},\mC_{B}) \to \mC'=(\mC_{A}',\mC_{B}')$ involving an exchange of particle between $A$ and $B$, and its time-reversed counterpart reads
\begin{equation}
  \label{eq:DB_violation_micro_contact}
  \lim_{\ddr t \to 0} \frac{ P(\mC', t +\ddr t ; \mC, t) }{P(\mC, t + \ddr t ; \mC', t)} = \frac{P(\mC) T_{c}(\mC'|\mC)}{P(\mC') T_{c}(\mC|\mC')} \neq 1 
\end{equation}
since $P(\mC) \neq P_{eq}(\mC)$ for generic non-equilibrium steady states.

\subsubsection{Analysis in terms of force and mobility}
\label{sec:force_and_activity_transrate}

Each transition rate $\pi(\rho_{A}'|\rho_{A})$ can indeed be decomposed as
\begin{equation}
  \label{eq:decomposition_transrates_force_activity}
  \pi(\rho_{A}'|\rho_{A}) = a_{V_{A}}(\rho_{A},\Delta N_{A}) e^{\frac{1}{2} F_{V_{A}}(\rho_{A},\Delta N_{A})}
\end{equation}
where, by definition,
\begin{equation}
  \label{eq:def:force-activity}
  \eqalign{
  F_{V_{A}}(\rho_{A},\Delta N_{A}) & = \ln \frac{\pi(\rho_{A}'|\rho_{A})}{\pi(\rho_{A}|\rho_{A}')} \\
  a_{V_{A}}(\rho_{A},\Delta N_{A}) & = \sqrt{\pi(\rho_{A}'|\rho_{A})\pi(\rho_{A}|\rho_{A}')}  \, .
  }
\end{equation}
$F_{V_{A}}(\rho_{A}, \Delta N_{A})$ is interpreted a a \emph{thermodynamic force} and is anti-symmetric with respect to the transition $\rho_{A}\to\rho_{A}+\Delta N_{A}/V_{A}$: $F_{V_{A}}(\rho_{A},\Delta N_{A})=-F_{V_{A}}(\rho_{A}+\frac{\Delta N_{A}}{V_{A}}, - \Delta N_{A})$. As for the symmetric quantity $a_{V_{A}}(\rho_{A},\Delta N_{A})=a_{V_{A}}(\rho_{A}+\frac{\Delta N_{A}}{V_{A}}, -\Delta N_{A})$, it is sometimes referred to as a \emph{mobility} \cite{maes2007and,kaiser2018canonical}. One should stress that this decomposition is only a more convenient and physically meaningful rewriting of the coarse-grained transition rates $\pi$.
In the thermodynamic limit, when $V\to \infty$, the expressions given in \eqref{eq:def:force-activity} converge to
\begin{eqnarray}
  \label{eq:force-activity_LD}
  F(\rho_{A},\Delta N_{A}) && = \ln \frac{\varphi(\rho_{A},\Delta N_{A})}{\varphi(\rho_{A}, -\Delta N_{A})} = - F(\rho_{A},-\Delta N_{A}) \\
  a(\rho_{A},\Delta N_{A}) && = \sqrt{\varphi(\rho_{A},\Delta N_{A}) \varphi(\rho_{A},-\Delta N_{A}) } = a(\rho_{A},-\Delta N_{A}) \; .      \nonumber 
\end{eqnarray}

Breaking of the macroscopic detailed balance \eqref{eq:macro_detailed_balance} can thus be re-expressed in terms of the thermodynamic force $F(\rho_{A}, \Delta N_{A})$. According to \eqref{eq:macro_detailed_balance} and \eqref{eq:force-activity_LD}, one can generally write $F = -I' \Delta N_{A} + F^{(A)}$ where $I'$ is here the general solution of the Hamilton Jacobi equation \eref{eq:hamilton-jacobi_eq}. Hence, using these new notation, macroscopic detailed balance simply corresponds to $F^{(A)}(\rho_{A},\Delta N_{A}) = 0$. This reformulation will be used in the sequel.

\subsection{Breaking of macroscopic detailed balance: perturbative expansion}
\label{sec:absence_macro_DB}

When macroscopic detailed balance does not hold, one has to come back to the complete Hamilton-Jacobi equation \eqref{eq:hamilton-jacobi_eq}
whose steady-state solution is $I(\rho_{A}|\bar{\rho})$. One has seen in Sec.~\ref{sec:macro_detailed_balance} that when only \emph{one} particle can be exchanged, the dynamics necessarily obeys macroscopic detailed balance. But if one can exchange more than one particle at a time through some cooperative mechanism at contact, macroscopic detailed balance may not hold. This is also the case when the exchanged quantity is continuous. An explicit example will be presented in Sec.~\ref{sec:MTM}.

For convenience, we use in this section the Hamilton-Jacobi equation formulated in terms of forces and activity (see section \ref{sec:force_and_activity_transrate} and equation \eqref{eq:recall_HJeq_F-a_Iss} below).
When two particles at most can be exchanged ($\Delta N_{A} = \pm 1, \, \pm 2$), the Hamilton-Jacobi equation is a fourth order polynomial equation in $e^{I'(\rho_{A}|\bar{\rho})}$, whose solution already takes a complicated form. 
For more than two particles exchanges as well as for a continuous exchanged quantities, there is no general way to compute exactly solutions of the Hamilton-Jacobi equation (\ref{eq:macroDB}). One may instead perform a perturbative expansion around some known reference solution, generally taken to be the solution verifying macroscopic detailed balance (equilibrium solution).

To perform a perturbative expansion, we follow the same lines as the one already expounded for diffusive systems in the weak noise limit \cite{bouchet2016perturbative}. 
We remain here at a formal level and do not discuss convergence issues regarding the expansions.
We call $\xi$ the parameter which characterises the ``distance'' between the solution $I^{\rm (DB)}(\rho_{A}) =I^{(0)}(\rho_{A})$ (we omit the $\bar{\rho}$-dependence here, to lighten notations) that verifies the macroscopic detailed balance and the one we are looking for, $I(\rho_{A})$. We further set
\begin{eqnarray}
  \label{eq:perturbation_expansion_F-a-Iss}
  I(\rho_{A}) && = \sum_{n \geqslant 0} \xi^{n} \, I^{(n)}(\rho_{A})  \\
  F(\rho_{A},\Delta N_{A})  && = \sum_{n\geqslant 0} \xi^{n} \, F^{(n)}(\rho_{A}, \Delta N_{A}) \\
  a(\rho_{A}, \Delta N_{A}) && = \sum_{n\geqslant 0} \xi^{n} \, a^{(n)}(\rho_{A},\Delta N_{A}) 
\end{eqnarray}
where $F^{(0)}(\rho_{A}, \Delta N_{A}) = - \Delta N_{A} I^{\prime}(\rho_{A})$ according to macroscopic detailed balance.

Since we would like to perform this perturbation expansion in terms of $F$, $a$ and $I(\rho_{A})$ -- which happens to be easier to handle --, we write here the Hamilton-Jacobi equation \eqref{eq:hamilton-jacobi_eq} in terms of the latter quantities:
\begin{equation}
  \label{eq:recall_HJeq_F-a_Iss}
\!\!\!\!\!\!\!\!\!\!\!\!\!\!\!\!\!\!\!\!\!\!\!\!\!\!\!\!\!\!
  \sum_{\Delta N_{A} > 0} \sinh\left(I^{\prime}(\rho_{A}) \Delta N_{A} \right) a(\rho_{A},\Delta N_{A}) \sinh \left(F(\rho_{A},\Delta N_{A}) + I^{\prime}(\rho_{A}) \Delta N_{A} \right) = 0 \, .
\end{equation}
Performing the expansion order by order up to $\Or\left(\xi^{2}\right)$ yields
\begin{itemize}
\item $\Or\left(\xi^{0}\right)$: 
  \begin{equation}
    \label{eq:order_0_exp_HJ}
    \fl
     I^{(0)\, \prime}(\rho_{A}) = - \frac{1}{\Delta N_{A}}F^{(0)}(\rho_{A},\Delta N_{A}) = F^{(0)}(\rho_{A}, -1)  \, ,
   \end{equation}
   according to detailed balance.
\item $\Or\left(\xi^{1}\right)$:
  \begin{equation}
    \label{eq:order_1_exp_HJ}
    \fl
    \eqalign{
    & I^{(1) \, \prime}(\rho_{A}) J^{(0)}(\rho_{A}) = \\
    & \qquad - \!\!\!\! \sum_{\Delta N_{A} > 0} a^{(0)}(\rho_{A}, \Delta N_{A}) \sinh\left(I^{(0)\, \prime}(\rho_{A}) \Delta N_{A} \right) F^{(1)}(\rho_{A}, \Delta N_{A})  \; ,
  }  
  \end{equation}
  with $J^{(0)}(\rho_{A})=2 \sum_{\Delta N_{A}} \Delta N_{A} \, a^{(0)}(\rho_{A},\Delta N_{A})\sinh(F^{(0)}(\rho_{A},\Delta N_{A}))$, the macroscopic current of the dynamics at order $\Or\left(\xi^{0}\right)$.
\item $\Or\left(\xi^{2}\right)$:
  \begin{equation}
    \label{eq:order_2_exp_HJ}
    \fl
    \eqalign{
    &    I^{(2) \, \prime}(\rho_{A}) J^{(0)}(\rho_{A}) =   \\
    & \quad   -  \sum_{\Delta N_{A} > 0} \!\!  a^{(0)}(\rho_{A}, \Delta N_{A}) \sinh\left(I^{(0) \, \prime}(\rho_{A}) \Delta N_{A} \right) F^{(2)}(\rho_{A}, \Delta N_{A})   \\
    & \quad  -  \sum_{\Delta N_{A} > 0} \left( F^{(1)}(\rho_{A},\Delta N_{A}) + I^{(1) \, \prime}(\rho_{A}) \Delta N_{A} \right)  \\
    & \qquad \quad  \times \left[ a^{(0)}(\rho_{A},\Delta N_{A}) \cosh\left(I^{(0)\,\prime}(\rho_{A}) \Delta N_{A} \right) I^{(1) \, \prime}(\rho_{A}) \Delta N_{A} \right.   \\
    & \qquad \qquad \quad    \left. \ + a^{(1)}(\rho_{A},\Delta N_{A}) \sinh\left(I^{(0)\, \prime}(\rho_{A})\Delta N_{A} \right) \right] \; .
    }
  \end{equation}
  
\end{itemize}
One can thus infer that the perturbative expansion may be formally rewritten as, for $k\geqslant 1$,
\begin{equation}
  \label{eq:scheme_perturbation_exp_HJ}
  I^{(k)\, \prime}(\rho_{A}) J^{(0)}(\rho_{A}) = \mathcal{F}^{(k)}\left[I^{(0)}, I^{(1)}, \dots{}, I^{(k-1)}\right]\left(\rho_{A} \right) \, .
\end{equation}
The contribution $I^{(k)}$ can thus be iteratively computed by integrating $I^{(k)\, \prime}$ along \jg{the trajectory $\rho_{A}(t)$ that follow the reference dynamics ($\xi=0$). For $\rho_{A}(t)$ such that $\dot{\rho}_{A}(t) = J^{(0)}(\rho_{A}(t))$, one has indeed}
\begin{equation}
\mathrm{d}I^{(k)}(\rho_{A}(t))= I^{(k)\, \prime}(\rho_{A}(t))J^{(0)}(\rho_{A}(t)) \mathrm{d}t 
\end{equation}
Performing the integration along the path starting from the stationary point $\rho_{A}^{\ast \, (0)}$ of the reference dynamics and ending at a point $\rho_{A}$ at $t\to\infty$ \jg{(called ``fluctuation path'', see~\ref{app:path_integral_LD})}, one obtains
\begin{equation}
  \label{eq:integration_perturbation_exp_relaxation_dyn}
  I^{(k)}(\rho_{A}) = C^{(k)} +  \int\limits_{0}^{\infty}  \mathrm{d}t \; \mathcal{F}^{(k)}\left[I^{(0)}, \dots{},I^{(k-1)} \right]\left(\rho_{A}(t) \right) \, ,
\end{equation}
where $C^{(k)} =  I^{(k)}(\rho_{A}^{\ast\, (0)})$ is an unknown constant to be determined. A simple way to compute this constant is to remember that the stationary point $\rho_{A}^{\ast\, (\xi)}$ of the perturbed dynamics verifies $I^{(\xi)}(\rho_{A}^{\ast\, (\xi)}) = 0$. The stationary point $\rho_{A}^{\ast \, (\xi)}$ itself can be computed by looking for the stationary solution of the macroscopic dynamics 
\begin{equation}
  \label{eq:stationary_solution_current_order_k} 
  J^{(\xi)}(\rho_{A}^{\ast\, (\xi)}) = 0 \, ,
\end{equation}
which allows one to look for $\rho_{A}^{\ast \, (\xi)}$ as $\rho_{A}^{\ast \, (\xi)} = \sum_{n\geqslant 0} \xi^{n} \rho_{A}^{\ast \, (n)}$. Then, Taylor expanding $J^{(\xi)}(\rho_{A}^{\ast \, (\xi)}) = 0$ allows us to express $C^{(k)} = I^{(k)}(\rho_{A}^{\ast \, (0)})$ as function of $\rho_{A}^{\ast \, (0)}$, the coefficients $\rho_{A}^{\ast \, (n)}$ of the Taylor expansion of $\rho_{A}^{\ast \, (\xi)}$, the functions $I^{(n)}(\rho_{A})$ and their derivatives for $n<k$.


Such a formal expansion has been outlined here to show how one may solve the general Hamilton-Jacobi equation when the dynamics of the density $\rho_{A}$ does not obey detailed balance. We will use the first order of this expansion in Sec.~\ref{sec:not_additive} when discussing the additivity property of the large deviations function $I$ in cases when macroscopic detailed balance does not hold.

\subsection{Non-additive large deviations function of the densities}
\label{sec:not_additive}

The large deviations function $I(\rho_A|\bar\rho)$ is a convenient tool to define non-equilibrium chemical potentials \cite{bertin2006def,bertin2007intensive,pradhan2010nonequilibrium,pradhan2011approximate,chatterjee2015zeroth,guioth2018large,guioth2019lack,GuiothTBP19}
As mentioned in the introduction, the definition of a non-equilibrium chemical potential relies on an additivity condition on the large deviations function, which reads \cite{bertin2006def,bertin2007intensive,guioth2018large,GuiothTBP19}
\begin{equation}
  \label{eq:additivity_LDF}
  \mathcal{I}(\rho_{A},\rho_{B}) \equiv \gamma_{A} I(\rho_{A}|\bar{\rho}) = \gamma_{A}I_{A}(\rho_{A}) + \gamma_{B}I_{B}(\rho_{B})
\end{equation}
where the density $\rho_B$ is given by $\rho_{B}= \gamma_{B}^{-1}(\rho-\gamma_{A}\rho_{A})$, with $\gamma_A=V_A/V$ and  $\gamma_B=V_B/V$.
Note that the functions $I_{A}(\rho_{A})$ and $I_{B}(\rho_{B})$ depend only on the density in one system, and not on the overall density $\bar\rho$.

\jg{A sufficient condition for the additivity property \eqref{eq:additivity_LDF} to hold} has been found in \cite{guioth2018large,GuiothTBP19}. It relies both on the macroscopic detailed balance condition \eqref{eq:macro_detailed_balance},
\begin{equation} \label{eq:LD_macro_detailed_balance}
  I^{\prime}(\rho_{A}) = \ln \frac{\varphi(\rho_{A}, -1)}{\varphi(\rho_{A}, +1)}\,  
\end{equation}
and on a factorization property of the coarse-grained transition rate at contact,
\begin{equation}
  \label{eq:factorization_transrate}
  \varphi(\rho_{A},\Delta N_{A}) = \nu_{0}\phi_{A}(\rho_{A},\Delta N_{A})\phi_{B}(\rho_{B},\Delta N_{B})
\end{equation}
with $\Delta N_{B} = -\Delta N_{A}$ and $\nu_{0}$ a frequency scale.

If the additivity condition \eqref{eq:additivity_LDF} is satisfied, the steady-state densities $\rho_{A}^{\ast}$ and $\rho_{B}^{\ast}$ are determined by the equality of the non-equilibrium chemical potentials $\mu_A=I_{A}'(\rho_{A}^{\ast})$ and $\mu_B=I_{B}'(\rho_{B}^{\ast})$. 
When the additivity condition \eqref{eq:additivity_LDF} is not satisfied, no chemical potential can be defined, and the steady-state densities have to be determined directly by minimizing the large deviations function $\mathcal{I}(\rho_{A},\rho_{B})$ under the constraint $\rho_{B}= \gamma_{B}^{-1}(\bar{\rho}-\gamma_{A}\rho_{A})$.
We discuss in this section two generic situations where the additivity property does not hold, and where the large deviations function $\mathcal{I}(\rho_{A},\rho_{B})$ can be determined explicitly, either exactly or through a perturbative expansion.

\subsubsection{Macroscopic detailed balance with non-factorized contact dynamics}
\label{sec:non-factorised_DB}

When the macroscopic detailed balance holds, the large deviations function of the density is explicitly given by \eqref{eq:LD_macro_detailed_balance} in terms of the coarse-grained rates $\varphi(\rho_{A},\pm 1)$.
However, in some cases the rates $\varphi(\rho_{A},\pm 1)$ cannot be written in the factorized form \eqref{eq:factorization_transrate}.
This lack of factorization at a coarse-grained level is in general due to a similar lack of factorization of the microscopic transition rate $T_{c}(\mC^{\prime}|\mC)$, as occurs for instance for the heat-bath rule or the Metropolis rule for instance \cite{GuiothTBP19}. The relation $I^{\prime}(\rho_{A}^{\ast}|\bar{\rho})=0$ can nevertheless still be used to determine the steady-state densities $\rho_{A}^{\ast}$ and $\rho_{B}^{\ast}$. 


\subsubsection{Absence of macroscopic detailed balance}

Now, if macroscopic detailed balance is broken, the Hamilton-Jacobi equation (\ref{eq:hamilton-jacobi_eq}) is much more difficult to solve to obtain $I(\rho_{A}|\bar{\rho})$.
One may however perform a perturbative expansion with respect to a reference state, assumed to satisfy the macroscopic detailed balance (see \sref{sec:absence_macro_DB} for details). To first order in the perturbation parameter $\xi$, the derivative of the large deviations function reads as
\begin{equation}
  \label{eq:LD_first_order_perturbation_nonadd}
  I'(\rho_{A}|\bar{\rho}) = I^{(0)\,\prime}(\rho_{A}|\bar{\rho}) + \xi I^{(1)\,\prime}(\rho_{A}|\bar{\rho}) + \Or\left(\xi^{2}\right) \; .
\end{equation}
\jg{In the specific case where the reference state is chosen as the equilibrium state, one has simply $I^{(0) \, \prime}=\mu_{A}^{\mathrm{eq}} - \mu_{B}^{\mathrm{eq}}$}. Let us assume now that the macroscopic transition rates completely factorise at each order in $\xi$. Contrary to what is expounded in \ref{sec:absence_macro_DB}, it is easier to formulate things in terms of transition rates $\varphi(\rho_{A},\Delta N_{A})$ directly. $I^{(1)\,\prime}(\rho_{A}|\bar{\rho})$ reads as
\begin{equation}
  \label{eq:I1_perturbation_cg_transrate}
  I^{(1)\, \prime}(\rho_{A}|\bar{\rho}) = \frac{\sum_{\Delta N_{A} \neq 0} \varphi^{(1)}(\rho_{A}, \Delta N_{A})\left( e^{I^{(0)\, \prime}(\rho_{A}|\bar{\rho})\Delta N_{A}} - 1  \right)}{\sum_{\Delta N_{A}\neq 0} \varphi^{(0)}(\rho_{A},\Delta N_{A}) \Delta N_{A}} \; .
\end{equation}
\jg{We point out that $I^{(1) \, \prime}$ does not vanishes at $\rho_{A}^{(0)}$ for which $J(\rho_{A}^{(0)}) = \sum_{\Delta N_{A} \neq 0} \varphi^{(0)}(\rho_{A}^{(0)}, \Delta N_{A}) \Delta N_{A} = 0$. This is because $I^{(0) \, \prime} (\rho_{A}^{(0)})=0$ as well, so that the ratio stays finite.}

Hence, even if $\varphi^{(1)}(\rho_{A},\Delta N_{A})$ and $\varphi^{(0)}(\rho_{A},\Delta N_{A})$ factorise as in (\ref{eq:factorization_transrate}), it is not expected that this factorisation property still holds for the large deviations function $I(\rho_{A}|\bar{\rho})$, even at the first order beyond the equilibrium situation for which the large deviations function is additive. Explicit examples of this lack of additivity when macroscopic detailed balance breaks down will be discussed in Sec.~\ref{sec:MTM}, in the framework of an exactly solvable lattice model.

Before that, we further discuss the thermodynamic relevance of the large deviations function $I(\rho_{A}|\bar{\rho})$ of the density, and explore in the next \sref{sec:potential} how one can physically measure the latter.


\section{Measure of $I(\rho_{A}|\bar{\rho})$: bias and generalized second law of thermodynamics}\label{sec:potential}


To go beyond formal calculations, an issue of practical importance would be to be able to measure the large deviations function $I(\rho_A|\bar{\rho})$.
Similarly to the measurement of free energies at equilibrium, one may think about the following approaches:
\begin{enumerate}
  \item A dynamical approach which consist of measuring the supplied work associated with a given protocol, starting from a known configuration of densities;
  \item A static approach that allows one to get access to the derivative of the large deviations function $I(\rho_A|\bar\rho)$ by biasing the dynamics (typically by applying external potentials) and measuring the stationary densities.
\end{enumerate}




\subsection{Static bias of the dynamics}\label{sec:bias_dyn}

A natural and interesting way to access the large deviations function $I$ (see \cite{sasa2006steady} for a phenomenological account), is to bias the dynamics to make rare events of the unbiased dynamics typical \cite{touchette2009large}. We would like to emphasise here the connections between a general bias $\lambda$ (which can be different from an external potential difference) and the large deviations function $I$, by using a more formal approach.
\jg{A direct and simple connection between a bias $\lambda$ and $I$ exists when the former is the conjugated variable of the density $\rho_{A}$ through a Legendre transform}. This can be seen by considering the stationary moment generating function
\begin{equation}
  \label{eq:generating_moment_function}
  \mathcal{Z}(\lambda) = \left\langle e^{\lambda V_{A} \rho_{A}} \right\rangle = \sum_{\rho_{A}} P(\rho_{A}|\bar{\rho})e^{\lambda V_{A} \rho_{A}} \; . 
\end{equation}
When $P(\rho_{A}|\bar{\rho}) \sim \exp\left(-V_{A} I(\rho_{A}|\bar{\rho}) \right)$ with $I$ the associated large deviations function, one has
\begin{equation}
  \label{eq:LD_gen_moment_function}
  \chi(\lambda) = \lim_{V_{A}\to \infty} \frac{1}{V_{A}} \ln \mathcal{Z}(\lambda) = \lambda \rho_{A}^{\ast}(\lambda) - I(\rho_{A}^{\ast}(\lambda)|\bar{\rho}) \, ,
\end{equation}
where $\rho_{A}^{\ast}(\lambda) = {\rm argmax}_{\rho_{A}}(\lambda \rho_{A} - I(\rho_{A}|\bar{\rho})))$ is the unique saddle-point (in the absence of any first order phase transition, \emph{i.e.} when $I$ is assumed to be convex) that satisfies 
\begin{equation}
  \label{eq:argmax_LD_gen_moment_function}
  \lambda = I^{\prime}(\rho_{A}^{\ast}(\lambda)|\bar{\rho}) \, .
\end{equation}
As a consequence, if one knows the bias $\lambda$ and is able to measure $\rho_{A}^{\ast}$ by another mean, the value of the derivative of the large deviations function is obtained from (\ref{eq:argmax_LD_gen_moment_function}). Yet, Eq.~(\ref{eq:argmax_LD_gen_moment_function}) is purely formal and essentially of no use if the tilt $\lambda V_{A} \rho_{A}$ cannot be implemented in practice.

\jg{In order to implement such a tilt, an experimenter will only be able to modify the dynamics, \emph{i.e.} the transition rates.}
To discuss this issue in a general framework, we consider macroscopic transition rates $\varphi(\lambda ; \rho_{A}, \Delta N_{A})$ that depend on an extra control parameter $\lambda$. The parameter $\lambda$ is generic and may be an external potential difference applied between the two systems, or even the driving force applied on one or both systems, etc. We choose $\lambda$ such that $\varphi( \lambda=0 ; \rho_{A}, \Delta N_{A}) = \varphi(\rho_{A} , \Delta N_{A})$. 

To determine the dependence of the large deviations function $I(\lambda; \rho_A|\bar\rho)$ on $\lambda$, we evaluate the derivative of $I(\lambda; \rho_A|\bar\rho)$ with respect to $\lambda$. \jg{The tilt $\lambda$ acts linearly as in (\ref{eq:argmax_LD_gen_moment_function}) if $\partial I^{\prime}/\partial \lambda$ is a constant (independent of $\rho_{A}$ and $\lambda$). Indeed, $\partial I'/\partial \lambda = -K$ yields by integration $I'(\lambda ;  \rho_{A} |\bar{\rho}) = I'(\rho_{A}|\bar{\rho}) - K\lambda$. Evaluating the latter equality at $\rho_{A}=\rho_{A}^{\ast}(\lambda)$, the stationary points of the biased dynamics ($I'(\lambda;  \rho_{A}^{\ast}(\lambda) | \bar{\rho}) = 0$), one obtains $I'(\rho_{A}^{\ast}(\lambda))= K \lambda$ and hence \eref{eq:argmax_LD_gen_moment_function} by a rescaling of $\lambda$.}

Interestingly, the partial derivative $\partial I^{\prime}/\partial \lambda$ can be directly estimated from the Hamilton-Jacobi equation \eqref{eq:hamilton-jacobi_eq} associated with the modified dynamics
\begin{equation}
  \label{eq:HJ_equation_modified_dynamics}
  \sum_{\Delta N_{A} \neq 0} \varphi(\lambda ; \rho_{A} , \Delta N_{A}) \left( e^{I^{\prime}(\lambda ; \rho_{A}|\bar{\rho})\Delta N_{A}} - 1 \right) = 0 \, ,
\end{equation}
valid for all $\rho_{A}$.
Taking the derivative of \eqref{eq:HJ_equation_modified_dynamics} with respect to $\lambda$ gives
\begin{equation}
  \label{eq:derivative_modified_HJeq}
  \fl
  \pderiv{I^{\prime}}{\lambda} = \left( J^{\dag}(\lambda ; \rho_{A}) \right)^{-1} \left[\sum_{\Delta N_{A} \neq 0} \pderiv{\varphi}{\lambda}(\lambda ; \rho_{A} ; \Delta N_{A}) \left( e^{I^{\prime}(\lambda ; \rho_{A}|\bar{\rho}) \Delta N_{A}} - 1 \right) \right] \; ,
\end{equation}
with $J^{\dag}(\lambda;\rho_{A}) = \sum_{\Delta N_{A}\neq 0} \varphi(\lambda; \rho_{A},\Delta N_{A}) e^{I'(\lambda;\rho_{A}) \Delta N_{A}}$ the macroscopic current associated with the adjoint dynamics \cite[Appendix 1]{kipnis2013scaling} \jg{(see \ref{app:path_integral_LD} of this paper as well)}.
\begin{sloppypar}
The right hand side of the last equation is not constant in general.  
When $\varphi(\lambda ; \rho_{A} , \Delta N_{A})$ takes the form
$\varphi(\lambda ; \rho_{A} , \Delta N_{A}) = \varphi(\rho_{A} , \Delta N_{A})e^{\frac{\lambda}{2} \Delta N_{A}}$, 
equation \eref{eq:derivative_modified_HJeq} reads as
\end{sloppypar}
\begin{equation}
  \label{eq:derivative_modified_HJeq_simple_tilt}
  \pderiv{I^{\prime}}{\lambda} = - \frac{J(\lambda ; \rho_{A}) + J^{\dag}(\lambda ; \rho_{A})}{2 \, J^{\dag}(\lambda ; \rho_{A})} \; .
\end{equation}
which in general depends on $\lambda$.
When macroscopic detailed balance holds, $J=J^{\dag}$ (see \sref{sec:force_and_activity_transrate} and \ref{app:path_integral_LD}) and $\partial I^{\prime}/\partial \lambda = -1$, which yields \eref{eq:argmax_LD_gen_moment_function} \jg{as stated above}.

To summarize, the variation of the large deviations function $I(\lambda; \rho_A|\bar\rho)$ with the external parameter $\lambda$ is generically non-linear in $\lambda$. Even when the perturbation only changes the ``force'' $F(\rho ; \Delta N_{A})$ and not the ``mobility'' in the expression (\ref{eq:decomposition_transrates_force_activity}) of the macroscopic transition rate, the dependence of $I(\lambda; \rho_A|\bar\rho)$ on $\lambda$ may still take a complicated form. A sufficient condition for the perturbation of $I(\rho_A|\bar\rho)$ to be linear with respect to $\lambda$ is when macroscopic detailed balance holds and when the perturbation acts only on the force $F$ associated with the transition rate $\varphi$. This linear perturbation may be expected when the perturbation $\lambda=\Delta U$ is due to the presence of uniform external potentials, but measuring the derivative of the large deviations through the perturbation $\lambda$ may not be an easy task and generally requires additional information on the derivative $\partial I^{\prime}/\partial \lambda$ (see \eqref{eq:derivative_modified_HJeq}).

\subsection{Applied external potential difference and second law of thermodynamics}\label{sec:apply:potential}

\subsubsection{Influence of an external potential difference on the dynamics at contact}

\jg{In order to go beyond the static analysis of the previous section, we now investigate thermodynamic transformations that involve a protocol performed by an external operator. In order to fix a precise physical setting, we introduce external potentials $U_{A}$ and $U_{B}$ which are assumed uniform over each sub-systems $A$ and $B$ respectively. These external potentials thus bias the contact dynamics only.}


The aim of the present section is to show that, similarly to other analysis for diffusing systems at hydrodynamic space-time scale \cite{bertini2002macroscopic,bertini2015macroscopic}, the large deviations function $I(\rho_{A}|\bar{\rho})$ obeys a (generalized) second law under some conditions. 
In that respect, the large deviations function $I(\rho_{A}|\bar{\rho})$ plays the role of a (normalized) non-equilibrium free energy.

Let us first recall the situation at equilibrium. The balance of mass between two systems in contact, controlled by the chemical potentials of each system, can be biased by uniform external potentials. Adding a potential $U(x)=U_{A}$ for all $x\in\Lambda_{A}$ and $U(x)=U_{B}$ for all $x\in\Lambda_{B}$ leads to a balance of mass governed by the equalisation of generalised chemical potentials --- sometimes called \emph{electro-chemical potentials} --- $\mu_{k}^{\mathrm{eq}} + U_{k}$, $k=A,\,B$. When an operator changes $\Delta U = U_{A} - U_{B}$ over time, the latter applies a work on the system which brings the densities to a new equilibrium at the end of the transformation. If the transformation is quasi-static, i.e., the transformation speed is infinitely slow, then the average supplied work is equal to the difference in the free energy of the whole system. Otherwise, the second law states that the average supplied work is greater than the difference of free energies between the initial and final states.

Coming back to the non-equilibrium situation, we need to include explicitly the potential energy difference $\Delta U$ in the microscopic transition rates at contact, $T_{c}^{\Delta U}$. We naturally assume that tilted transition rates at contact satisfy local detailed balance (\ref{eq:local_DB}) that reads here as
\begin{equation}
  \label{eq:tilted_transrate}
  \fl
  \frac{T_{c}^{\Delta U}(\mC_{A}', \mC_{B}'|\mC_{A},\mC_{B})}{T_{c}^{\Delta U}(\mC_{A}, \mC_{B}|\mC_{A}',\mC_{B}')} = e^{-\beta \left[ \left( E_{A}(\mC_{A}')-E_{A}(\mC_{A}) + E_{B}(\mC_{B}')-E_{B}(\mC_{B}) \right)  + \Delta U \left(\mathcal{N}(\mC_{A}')-\mathcal{N}(\mC_{A}) \right)  \right]} \; ,
\end{equation}
with $\mathcal{N}(\mC_{A})$ the total particle number in system $A$ in the configuration $\mC_{A}$.


We now specify the protocol that the operator performs to go from the initial value $\Delta U_{i}$ and $\Delta U_{f}$. For all $s\in [0,1]$, one defines a function $\Delta U(s)$ such that $\Delta U_{0}=\Delta U_{i}$ and $\Delta U_{1}=\Delta U_{f}$. Assuming that the total real time to apply this transformation is $T$, we interpret the parameter $s$ as $s=\tau/T$, where we recall that $\tau$ is the re-scaled macroscopic time $\tau=\epsilon t$ introduced in Sec.~\ref{sec:multi-scale_analysis}. Importantly, we assume that the rate of change of the potential difference $\Delta U(\tau/T)$ is sufficiently low so that the time scale separation between the internal (bulk) dynamics of both systems and the dynamics at contact still holds (see \sref{sec:multi-scale_analysis}). Along with this hypothesis, the time-dependent macroscopic transition rate $\varphi_{ \tau}^{\Delta U}$ hence reads as
\begin{equation}
  \label{eq:def:macro_trans_rate_protocol}
  \varphi_{\tau}^{\Delta U} = \varphi(\Delta U(\case{\tau}{T}) ; \rho, \Delta N)
\end{equation}
for all $\tau \in [0, T]$. The biased transition rate $\varphi(\Delta U, \rho, \Delta N)$ is an average of the tilted microscopic transition rate $T_{c}^{(\Delta U)}$ with respect to the stationary microscopic distribution of the isolated systems, as in (\ref{eq:detailed_coarse-grained_transrate_contact}).
As mentioned above, since external potentials are uniform over each system, the bulk transition rates are not perturbed by these potentials, and the stationary distributions of the isolated systems $P_{k}(\mC_{k}|\rho_{k})$ are the same as in the absence of any external potential. The coarse-grained transition rate at fixed potential difference $\Delta U$ thus reads as
\begin{equation}
  \label{eq:coarse_grained_tilted_extpot_transrate}
  \fl
  \varphi^{\Delta U}(\rho_{A},\Delta N_{A}) = \hat{\sum}^{(\Delta N_A)} \hat{\sum}^{(0)} T_{c}^{\Delta U}( \mC_{A}',\mC_{B}'| \mC_{A}, \mC_{B}) P_{V_{A}}(\mC_{A}|\rho_{A}) P_{V_{B}}(\mC_{B}|\rho_{B}) \, ,
\end{equation}
where the notation $\hat{\sum}^{(\Delta N_A)}$ has been defined in Eq.~\eqref{eq:detailed_coarse-grained_transrate_contact}.

A key issue is how the coarse-grained transition rate $\varphi^{\Delta U}$, and especially its associated force $F^{\Delta U}(\rho_{A},\Delta N_{A})$ (\ref{eq:force-activity_LD})
\begin{equation}
  \label{eq:force_cg_transrate_extpot_tilted}
  F^{\Delta U}(\rho_{A},\Delta N_{A}) = F(\Delta U ; \rho_{A},\Delta N_{A}) \equiv \ln \frac{ \varphi^{\Delta U}(\rho_{A}, \Delta N_{A})  }{ \varphi^{\Delta U}(\rho_{A}, - \Delta N_{A})} 
\end{equation}
depend on the external potential difference. At equilibrium, $F^{\Delta U}(\rho_{A},\Delta N_{A})=-\left(\mu_{A}^{\mathrm{eq}}(\rho_{A}) - \mu_{B}^{\mathrm{eq}}(\rho_{B}) + \Delta U \right) \Delta N_{A}$ and thus, the force would be linear in $\Delta U$.
But out-of-equilibrium, the force may not be linear.
As can be seen in the microscopic definition of the tilted coarse-grained transition rate (\ref{eq:coarse_grained_tilted_extpot_transrate}), this is a natural consequence of the choice of $T_{c}^{\Delta U}$.
This leads us to consider two different classes of transition rates.
\begin{enumerate}
\item The first one is the one for which the coarse-grained generalised force (\ref{eq:force_cg_transrate_extpot_tilted}) is linear in $\Delta U$. Hence, the coarse-grained transition rate reads
  \begin{equation}
    \label{eq:macro_transrate_tilt_extpot_lin}
    \varphi(\Delta U ; \rho_{A}, \Delta N_{A}) = a^{\Delta U}(\rho_{A},\Delta N_{A}) e^{\frac{1}{2}\left( F(\rho_{A},\Delta N_{A})  -  \beta \Delta U \Delta N_{A} \right)} \, ,    
  \end{equation}
  with $F(\rho_{A}, \Delta N_{A})$ the force in the absence of external potentials ($\Delta U=0$). 
\item The second one allows for a more complex dependence in $\Delta U$. It reads, in full generality, as 
  \begin{equation}
    \label{eq:macro_transrate_tilt_expot_gen}
    \varphi(\Delta U ; \rho_{A}, \Delta N_{A}) = a^{\Delta U}(\rho_{A}, \Delta N_{A}) e^{\frac{1}{2} F(\Delta U ; \rho_{A}, \Delta N_{A}) } \; ,
  \end{equation}
  where the force $F(\Delta U ; \rho_{A}, \Delta N_{A})$ is non-linear in $\Delta U$. This is for instance the case for other microscopic rules, different from the Sasa-Tasaki and the exponential ones, such as the Metropolis rule or the heat-bath (or Kawasaki) rule. 
\end{enumerate}

In both classes, one eventually introduces the macroscopic current $J(\Delta U, \rho)$ associated with the dynamics in presence of a bias $\Delta U$. It reads
\begin{equation}
  \label{eq:def:macro_current_with_bias}
  J(\Delta U, \rho_{A}) = \sum_{\Delta N_{A}} \Delta N_{A} \, \varphi(\Delta U ;  \rho_{A}, \Delta N_{A}) \; .
\end{equation}
The most probable relaxation path (which coincides with the average one at the thermodynamic limit) in the presence of a time-dependent forcing thus follows the relaxation dynamics
\begin{equation}
  \label{eq:relaxation_dynamics_with_tilt}
  \deriv{\rho_{A}(\tau)}{\tau} = J(\Delta U(\case{\tau}{T}) ; \rho_{A}(\tau)) \; .
\end{equation}

\subsubsection{Evaluation of the supplied work}

Notations being settled, one can enter the heart of this section which is the study of the work supplied by the protocol $\Delta U(s)$ for $s \in [0,1]$. The infinitesimal (averaged) work supplied by the external operator with the potential difference $\Delta U(s)$ between a time $\tau$ and $\tau + \ddr \tau$ reads as
\begin{equation}
  \label{eq:average_infinitesimal_work}
  \fl
  \eqalign{
  \left< \widehat{\delta W}\right>_{\tau} & = \delta W_{\tau} \\
  & = - \!\!\! \sum_{\Delta N_{A}\neq 0}\sum_{\rho_{A}} \Delta U(\case{\tau}{T}) \Delta N_{A} \varphi(\Delta U(\case{\tau}{T}) ; \rho_{A}, \Delta N_{A}) P_{\tau}^{\Delta U}( \rho_{A}|\bar{\rho}) \ddr \tau \, .
  }
\end{equation}
At large deviations level, Eq.~(\ref{eq:average_infinitesimal_work}) leads to the following integrated work
\begin{equation}
  \label{eq:average_work_thermolimit}
  \fl
  W_{[0,T]} = \int_{0}^{T} \!\! \delta W_{\tau} = - \int_{0}^{T} \!\! \sum_{\Delta N_{A} \neq 0} \!\! \Delta U(\case{\tau}{T}) \Delta N_{A} \varphi(\Delta U(\case{\tau}{T}) ; \rho_{A}(\tau) , \Delta N_{A}) \ddr \tau \, ,
\end{equation}
where $\rho_{A}(\tau)$ is the solution of the macroscopic equation \eref{eq:relaxation_dynamics_with_tilt}, as well as the minimum (saddle-point) of the large deviations function $I_{\tau}(\Delta U(\case{\tau}{T}), \rho_{A})$ associated with the distribution $P_{\tau}^{\Delta U}( \rho_{A})$ in the large volume limit $V\to\infty$.

For the sake of clarity when one will consider the quasi-static limit $T\to\infty$, it is convenient to change the integration variable from $\tau$ to $s=\tau/T$ that leads to
\begin{equation}
  \label{eq:average_work_thermolimit_adim_time}
  W_{[0,T]} = - T \int_{0}^{1} \!\! \ddr s \!\! \sum_{\Delta N_{A} \neq 0} \!\! \Delta U(s) \Delta N_{A} \varphi(\Delta U(s) ; \rho_{A}^{(T)}(s) ; \Delta N_{A}) \, ,
\end{equation}
where one has introduced $\rho_{A}^{(T)}(s)=\rho_{A}(sT)$.

Our aim is now to try to relate the supplied work $W_{[0,T]}$ to the large deviations function $I(\rho_{A}|\bar{\rho})$ of the unperturbed dynamics. 
One may achieve this goal by following the strategy developed by \cite{bertini2013clausius,bertini2012thermodynamic,bertini2015macroscopic,bertini2015quantitative} in the context of the Macroscopic Fluctuation Theory which itself dates back to the pioneering works of Oono \& Paniconi \cite{oono1998steady}, later developed by Sasa \& Hatano and Sasa \& Tasaki \cite{hatano2001steady,sasa2006steady}. 

In the present setting, the main idea is to replace $\Delta U(s) \Delta N_{A}$ in the expression (\ref{eq:average_work_thermolimit_adim_time}) of the work by the generalized forces $F(\rho_{A},\Delta N_{A})$ and $F(\Delta U ; \rho_{A}, \Delta N_{A})$. As previously discussed, $F(\Delta U ; \rho_{A}, \Delta N_{A})$ may be either linear in $\Delta U$ or not. It is thus convenient to isolate this linear dependence by introducing
\begin{equation}
  \label{eq:nonlinear_force_DU}
  F_{\mathrm{nl}}(\Delta U ; \rho_{A}, \Delta N_{A}) = F(\Delta U ; \rho_{A}, \Delta N_{A}) + \beta \Delta U \Delta N_{A} \; .
\end{equation}
where the index `nl' stands for `non-linear', meaning that $F_{\mathrm{nl}}(\Delta U ; \rho_{A}, \Delta N_{A})$ contains only non-linear contributions in $\Delta U$.

The two classes of contact dynamics introduced in Eqs.~(\ref{eq:macro_transrate_tilt_extpot_lin}) and (\ref{eq:macro_transrate_tilt_expot_gen}) thus differ in the $\Delta U$-dependence of $F_{\mathrm{nl}}(\Delta U;\rho_{A},\Delta{N_{A}})$: for the first class, $F_{\mathrm{nl}}(\Delta U;\rho_{A},\Delta N_{A})$ is independent of $\Delta U$ and is simply equal to $F(\rho_{A},\Delta N_{A})$, the force associated with the unbiased transition rates. In the second class, $F_{\mathrm{nl}}(\Delta U; \rho_{A},\Delta N_{A})$ does depend on $\Delta U$, and thus contains a genuine non-linearity in $\Delta U$.

In both cases, one can write $-\Delta U \Delta N_{A} = F(\Delta U ; \rho_{A}, \Delta N_{A}) - F_{\mathrm{nl}}(\Delta U ; \rho_{A}, \Delta N_{A})$ and the supplied work (\ref{eq:average_work_thermolimit_adim_time}) then reads
\begin{equation}
  \label{eq:average_work_general_forces}
  \fl
  \eqalign{
  \beta W_{[0,T]} & = T \int_{0}^{1}\ddr s \, \sum_{\Delta N_{A} \neq 0} \left[ F(\Delta U(s) ; \rho_{A}^{(T)}(s) , \Delta N_{A}) \right. \\ 
& \quad \left. -  F_{\mathrm{nl}}(\Delta U(s) ; \rho_{A}^{(T)}(s), \Delta N_{A})\right] 
\varphi(\Delta U(s) ; \rho_{A}^{(T)}(s) ; \Delta N_{A}) \, .
}
\end{equation}

The simpler case one can encounter is when both the property $F_{\mathrm{nl}}(\Delta U; \rho_{A},\Delta N_{A}=F(\rho_{A},\Delta N_{A})$ and the macroscopic detailed balance hold. One thus expects $F_{\mathrm{nl}}$ to be equal to $F(\rho_{A},\Delta N_{A})=-I'(\rho_{A}|\bar{\rho})\Delta N_{A}$, which already suggests a connection between the work supplied by the external potential and the large deviations function of the dynamics in the absence of external potential.
Hence, we first consider this simple case for which a second law holds before briefly moving toward the more complex case for which $F_{\mathrm{nl}}$ is genuinely non-linear in $\Delta U$, and thus different from the unperturbed dynamic force $F$.

\subsection{External potential difference as a linear bias in the forces}\label{sec:sub:first_case_work_principle}


We start with the first type of dynamics according to which $F_{\mathrm{nl}}(\Delta U ; \rho_{A}, \Delta N_{A}) = F(\rho_{A},\Delta N_{A})$ is independent of $\Delta U$. The average work then reads
\begin{eqnarray}
  \label{eq:average_work_general_forces}
\!\!\!\!\!\!\!\!\!\!\!\!\!\!\!\!
  \beta W_{[0,T]} && = T \int_{0}^{1}\ddr s \, \sum_{\Delta N_{A} \neq 0} \left[ F(\Delta U(s) ; \rho_{A}^{(T)}(s) , \Delta N_{A}) \right. \\ \nonumber
&& \qquad \qquad \qquad \quad \left. -  F( \rho_{A}^{(T)}(s), \Delta N_{A}) \right] 
\varphi(\Delta U(s) ; \rho_{A}^{(T)}(s) ; \Delta N_{A}) \, .
\end{eqnarray}
To establish a connection between the work and the large deviations function, it is useful to decompose the force $F$ into \cite{bertini2015macroscopic}
\begin{equation}
  \label{eq:sym_antisym_decomposition_forces}
  F(\rho_{A} , \Delta N_{A})  = -I'(\rho_{A}|\bar{\rho}) \Delta N_{A} +  F^{(A)}(\rho_{A}, \Delta N_{A})  \, ,
\end{equation}
where $F^{(A)}$ is responsible of the breaking of the time-reversible symmetry of the coarse-grained dynamics (see \ref{sec:sub:time-reversal}).
As for the time-dependent force $F(\Delta U(s); \rho_{A}^{(T)}(s), \Delta N_{A})$, one can introduce the associated decomposition
\begin{equation}\label{eq:sym_antisym_decomposition_forces_tilt_extpot}
  \fl
  F(\Delta U; \rho_{A}, \Delta N_{A}) = -I'(\Delta U, \rho_{A}|\bar{\rho}) \Delta N_{A} + F^{(A)}(\Delta U; \rho_{A}, \Delta N_{A}) \; ,
\end{equation}
where $I(\Delta U, \rho_{A} |\bar{\rho})$ stands for the stationary large deviations function associated with the contact dynamics at a \emph{fixed} external potential difference $\Delta U$.


Using the above decomposition, the work (\ref{eq:average_work_general_forces}) now reads
\begin{equation}
  \label{eq:average_work_forces_adim_time}
  \fl
  \eqalign{
  \beta W_{[0,T]}  & = T \int_{0}^{1} \ddr s \left( I^{\prime}(\rho_{A}^{(T)}(s)|\bar{\rho}) - I_{\Delta U(s)}^{\prime}(\rho_{A}^{(T)}(s)|\bar{\rho}) \right) J(\Delta U(s) ; \rho^{(T)}(s)) \\
                   & \quad  + T\int_{0}^{1} \ddr s \!\!\! \sum_{\Delta N_{A} \neq 0} \! \left[ F^{(A)}(\Delta U(s) ; \rho_{A}^{(T)}(s), \Delta N_{A}) - F^{(A)}(\rho_{A}^{(T)}(s) , \Delta N_{A}) \right]  \\
                   &  \qquad \qquad \qquad \qquad \qquad \qquad \times \varphi(\Delta U(s) ; \rho_{A}^{(T)}(s), \Delta N_{A}) \; .
                 }
\end{equation}
As seen in \sref{sec:force_and_activity_transrate}, the presence of the second term on the right hand side of Eq.~(\ref{eq:average_work_forces_adim_time}) is related to the breaking of time-reversal symmetry ---or, to put it another way, to the non-vanishing entropy production.
Hence this second term disappears if macroscopic detailed balance holds.

The first term under the integral in Eq.~(\ref{eq:average_work_forces_adim_time}) is a total derivative in time since $\rho_{A}^{(T)}(s)$ obeys Eq.~(\ref{eq:relaxation_dynamics_with_tilt}). The second term is however not a total time derivative because $I_{\Delta U(s)}'$ depends on time. The term in which $I_{\Delta U}'$ is involved can be seen as a macroscopic `free energy dissipation rate' (see \cite[eq. (26)]{ge2017mathematical}):
\begin{equation}
  \label{eq:def:free_energy_dissipation_rate}
  \dot{\mathcal{F}}_{\rm diss}(\Delta U ; \rho_{A}) \equiv - I_{\Delta U}'(\rho_{A}|\bar{\rho}) J(\Delta U ; \rho_{A}) \, .
\end{equation}
which turns out to be always positive \cite{ge2017mathematical}. \jg{This can be seen easily by remembering that $J(\Delta U ; \rho_{A}) = \sum_{\Delta N_{A}} \Delta N_{A} \varphi(\Delta U; \rho_{A}, \Delta N_{A})$ and by using the convex inequality $x \geqslant 1 - e^{-x}$ for $x=-I_{\Delta U}'(\rho_{A}|\bar{\rho}) \Delta N_{A}$. One thus finds that $\dot{\mathcal{F}}_{\rm diss} \geqslant \sum_{\Delta N_{A}} \varphi(\Delta U; \rho_{A}, \Delta N_{A})\left( \exp\left(\Delta N_{A}I_{\Delta U}'(\rho_{A}|\bar{\rho})\right) - 1 \right) = 0$.}

Finally, the first term in (\ref{eq:average_work_forces_adim_time}) reads as
\begin{equation}
  \label{eq:first_term_average_work_forces_admi_time}
  \left[I(\rho_{A}^{(T)}(1)|\bar{\rho}) - I(\rho_{A}^{(T)}(0)|\bar{\rho}) \right] + T \int_{0}^{1}\!\! \ddr s \, \dot{\mathcal{F}}_{\rm diss}(\Delta U(s), \rho_{A}^{(T)}(s)) \; .
\end{equation}

\subsubsection{When macroscopic detailed balance holds ($F^{(A)}=0$)}

As a result, when macroscopic detailed balance is obeyed, an analog of the second law of thermodynamics holds.
Indeed, the condition $F^{A}=0$ implies
\begin{equation}\label{eq:second_law_macro_DB_1}
  \fl
  \beta W_{[0,T]} = \left[I(\rho_{A}^{(T)}(1)|\bar{\rho}) - I(\rho_{A}^{(T)}(0)|\bar{\rho}) \right] + T \int_{0}^{1}\!\! \ddr s \, \dot{\mathcal{F}}_{\rm diss}(\Delta U(s), \rho_{A}^{(T)}(s)) \;
\end{equation}
which leads to
\begin{equation}\label{eq:second_law_macro_DB_2}
  \beta W_{[0,T]} \geqslant \left[I(\rho_{A}^{(T)}(1)|\bar{\rho}) - I(\rho_{A}^{(T)}(0)|\bar{\rho}) \right]   
\end{equation}
since the last term of \eref{eq:second_law_macro_DB_1} is positive.
One may intuitively expect that the last term in Eq.~(\ref{eq:second_law_macro_DB_1}) vanishes in the quasi-static limit $T\to\infty$. This result is shown more rigorously in \ref{sec:app:quasistatic}.

As a by-product of our analysis, we have derived, when detailed balance holds, that 
\begin{equation}
  \label{eq:link_large_dev_tilt_and_large_dev_no_tilt}
  I_{\Delta U}^{\prime}( \rho_{A}|\bar{\rho}) = I^{\prime}(\rho_{A}|\bar{\rho}) + \beta \Delta U \, .
\end{equation}
This result comes from the property $F^{(A)}=0$ (since macroscopic detailed balance holds), using the decomposition of the force $F$ given in Eqs.~(\ref{eq:sym_antisym_decomposition_forces}) and (\ref{eq:sym_antisym_decomposition_forces_tilt_extpot}), as well as the hypothesis that $F(\Delta U;\rho_{A},\Delta N_{A})$ is linear in $\Delta U$. 
Evaluating Eq.~(\ref{eq:link_large_dev_tilt_and_large_dev_no_tilt}) at $\rho_{A}=\rho_{A}^{\ast\, \Delta U}$ (\emph{i.e.} at the stationary state of the dynamics in the presence of a fixed bias $\Delta U$) cancels $I_{\Delta U}(\rho_{A}|\bar{\rho})$ and one obtains
\begin{equation}
  \label{eq:ST_relation_DeltaU_large_dev}
  I^{\prime}(\rho_{A}^{\ast \, \Delta U}|\bar{\rho}) + \beta\Delta U = 0 \, .
\end{equation}
If, in addition to macroscopic detailed balance, additivity holds, \eref{eq:ST_relation_DeltaU_large_dev} implies the relation postulated on phenomenological grounds by Sasa and Tasaki in their seminal paper \cite{sasa2006steady} that relates chemical potentials at contact and external potentials
\begin{equation}
  \label{eq:ST_relation_DeltaU_chempot}
  \mu_{A}^{\mathrm{cont}}(\rho_{A}^{\ast \, \Delta U}) + \beta U_{A} = \mu_{B}^{\mathrm{cont}}(\rho_{B}^{\ast \, \Delta U}) + \beta U_{B} \, .
\end{equation}
This relation can be used to measure the chemical potentials of, say, system $A$, using for system $B$ a reference system whose properties are known
\cite{sasa2006steady,GuiothTBP19}.

\subsubsection{When macroscopic detailed balance does not hold ($F_{A}\neq 0$)}

When macroscopic detailed balance does not hold, the supplied work $W_{[0,T]}$ (\ref{eq:average_work_forces_adim_time}) involves a house-keeping contribution \cite{oono1998steady,hatano2001steady,ge2017mathematical} that does not vanish, even in the quasi-static limit. This contribution reads, according to Eq.~(\ref{eq:average_work_forces_adim_time}),
\begin{equation}
  \label{eq:hk_term_work}
  \fl
  \eqalign{
   W_{[0,T]}^{\rm HK} &= T\int_{0}^{1} \!\!\! \ddr s \! \! \sum_{\Delta N_{A} \neq 0} \! \left[ F^{(A)}(\Delta U(s) ; \rho_{A}^{(T)}(s), \Delta N_{A}) \right. \\
& \qquad  \left. - F^{(A)}(\rho_{A}^{(T)}(s) , \Delta N_{A}) \right]
\varphi(\Delta U(s) ; \rho_{A}^{(T)}(s), \Delta N_{A}) \; .
}
\nonumber
\end{equation}
One might be tempted to establish a relation between $F^{(A)}(\Delta U ; \rho_{A}, \Delta N_{A})$ and $F^{(A)}( \rho_{A}, \Delta N_{A})$ when $F$ is linear in $\Delta U$.
However, this cannot be done since we do not know \emph{a priori} the dependence in $\Delta U$ of $I_{\Delta U}'$. The latter can be determined in general by the Hamilton-Jacobi equation (\ref{eq:hamilton-jacobi_eq}) of the tilted dynamics. But when macroscopic detailed balance is broken, the large deviations function may not be linear in $\Delta U$, even when $F(\Delta U ; \rho_{A}, \Delta N_{A})$ is linear in $\Delta U$ and when the mobility $a^{\Delta U}(\rho_{A},\Delta N_{A}) = a(\rho_{A},\Delta N_{A})$ does not depend on $\Delta U$.
This is confirmed by a perturbative expansion of the solution of the Hamilton-Jacobi (\ref{eq:hamilton-jacobi_eq}) for small $\Delta U$: the term of order $\Delta U^{2}$ does not vanish in general.

Finally, the house-keeping contribution (\ref{eq:hk_term_work}) cannot be avoided even in the quasi-static limit and has to be subtracted to evaluate the work $W_{[0,T]}$. If one knows by another mean the large deviations function, a measure of the work could be another possibility to assess the presence of a non-zero anti-symmetric force $F^{(A)}$ and thus the breaking of the macroscopic detailed balance.

\subsection{Non-linear bias in the forces}

When the force $F(\Delta U, \rho_{A},\Delta N_{A})$ is non-linear in $\Delta U$, one has to revert to the general relation \eref{eq:average_work_general_forces}. Since a general relation between $F_{\rm nl}$ and $F$ does not exist anymore, one cannot link the supplied work $W_{[0,T]}$ with the unbiased large deviations function $I$. As a consequence, any generalization of the second law for this non-linear $\Delta U$ dependence in the force $F(\Delta U; \rho_{A}, \Delta N_{A})$ does not seem to exist. A more involved analysis depending on the specific $\Delta U$ dependence of $F_{\rm nl}$ is thus required to assert any relationship between the work and the unbiased system.

\section{Application to an exactly solvable model}
\label{sec:MTM}

We now wish to illustrate the different results presented in the previous sections on the explicit example of an exactly solvable model, namely the driven lattice gas model introduced in Ref.~\cite{guioth2017mass}. We consider two different versions of the model, one with discrete particles moving on a lattice, and the other one with a continuous mass being exchanged between neighbouring sites.

\subsection{Definition of the driven lattice gas model}
\label{sec:def:MTM}


We start by defining the lattice gas model. We consider a one-dimensional lattice $\Lambda$, with an even number $|\Lambda|=2L$ of sites.
The number $n_{i}$ of particle on a site $i$ is at most $n_{\mathrm{max}}$.
This model draws inspiration from both an equilibrium kinetically constrained model \cite{PhysRevLett.95.015702} and from driven lattice gases like the Zero Range Process \cite{evans2005nonequilibrium}.
At odds with most standard driven lattice gas models, we assume the dynamics to be synchronous, and to involve two distinct partitions of the lattice, that we call $\mathcal{P}_{1}=\{(2k,2k+1)\}_{k\in\llbracket 0, L\rrbracket}$ and $\mathcal{P}_{2}=\{(2k+1,2k+2)\}_{k\in \llbracket 0, L \rrbracket}$, both of them gathering adjacent pairs of sites. At each step, one of the two partitions is randomly chosen with equal probability. Having selected a partition $\mathcal{P}_{j}$, all links of this partition are updated in parallel, in an independent way.
A link $(i,i+1)$ is updated according to the transition rate:
\begin{equation}
  \label{eq:def:transrate_MTM_model}
  \fl
  \eqalign{
  T(n_{i+1}^{\prime},n_{i}^{\prime}|n_{i+1},n_{i}) & = K(d_{i}^{\prime} | \bar{n}_{i}) \\
  & = \frac{\exp\left\{ -\left[\varepsilon\left(\bar{n}_{i}+\frac{d_{i}^{\prime}}{2}\right)+\varepsilon\left(\bar{n}_{i}-\frac{d_{i}^{\prime}}{2}\right)\right] + \frac{f}{2} d_{i}^{\prime} \right\}  }{Q(\bar{n}_{i})} .
  }
\end{equation}
with the notations $d_{i}^{\prime} = n_{i+1}^{\prime}-n_{i}^{\prime}$ and $\bar{n}_{i}=(n_{i}+n_{i+1})/2=(n_{i}^{\prime}+n_{i+1}^{\prime})/2$. The normalisation factor $Q(\bar{n}_{i})$ is such that $\sum_{n_{1}',n_{2}'} T(n_{1}', n_{2}'|n_{1},n_{2}) = 1$.
Conservation of the total particle number is enforced by setting the transition rate $T(n_{i+1}^{\prime},n_{i}^{\prime}|n_{i+1},n_{i})$ to zero if the conservation law $n_{i+1}^{\prime}+n_{i}^{\prime}= n_{i+1}+n_{i}$ is not satisfied. One must remark that the net transfer of particles $\Delta n_{i}=(n_{i}'-n_{i})=-(n_{i+1}'-n_{i+1})$ from site $i$ to site $i+1$ reads as $\Delta n_{i} = (d_{i}^{\prime}-d_{i})/2$ (with $d_{i}=n_{i+1}-n_{i}$). As a consequence, the probability to transfer $\Delta n_{i}$ particles according to the transition rate \eref{eq:def:transrate_MTM_model} is independent of the initial particle difference $d_{i}$ as one might expect intuitively for a mass transport model. 
The quantity $\varepsilon(n)$ plays the role of a local energy, while the parameter $f$ plays the role of a driving force. When $f=0$, the equilibrium Boltzmann-Gibbs distribution (with unit temperature) is recovered.
For nonzero $f$, the following local detailed balance relation holds
\begin{equation}
  \label{eq:ldb_MTM}
  \ln\frac{T(n_{i+1}^{\prime},n_{i}^{\prime}|n_{i+1},n_{i})}{T(n_{i+1},n_{i}|n_{i+1}^{\prime},n_{i}^{\prime})}=-\Delta \varepsilon_{i+1}-\Delta \varepsilon_{i} + f\Delta n_{i}
\end{equation}
where $\Delta \varepsilon_{i}=\varepsilon(n_{i}^{\prime})-\varepsilon(n_{i})$. 
The non-equilibrium steady-state distribution of the model is given by (see \cite{guioth2017mass})
\begin{equation}
  \label{eq:stationary_probability_MTM}
  P\left( \{ n_{i} \}_{i\in\Lambda} \right) = \frac{2}{Z(|\Lambda|,N)} \exp\left( \sum_{i\in\Lambda} \varepsilon(n_{i})\right) \cosh\left( \sum_{i\in \Lambda} (-1)^{i} f n_{i} \right) .
\end{equation}
One sees the driving force $f$ explicitly enters into the expression (\ref{eq:stationary_probability_MTM}) of the steady-state distribution, at odds with other exactly solvable models like the Zero Range Process \cite{evans2005nonequilibrium} or the Asymmetric Simple Exclusion Process on ring geometry.
This dependence on the drive is supposed to be generic \cite{mclennan1959statistical} and its occurrence here mostly justifies the interest of the present model.
From Eq.~(\ref{eq:stationary_probability_MTM}), the single site probability distribution $p(n|\rho)$ is evaluated as \cite[Appendix B]{GuiothTBP19}
\begin{equation}
\label{eq:single_site_probability_distribution_MTM}
p(n|\rho)= \frac{e^{\mu^{\rm iso}(\rho) - \varepsilon(n)}}{z_{0}(\mu^{\rm iso}(\rho))} \exp\left(\nu\left[\mu^{\rm iso}(\rho), f\right](n)\right)
\end{equation}
where one has introduced $z_{0}(x)=\sum_{n}e^{-\varepsilon(n) + xn}$, the normalisation constant for the equilibrium system at $f=0$; the non-equilibrium contribution
\begin{equation}\label{eq:noneq_term_single_site_proba_MTM}
  \exp\left(\nu\left[\mu, f\right](n)\right) = \frac{z_{0}(\mu)}{2} \left[ \frac{e^{fn}}{z_{+}(\mu)} + \frac{e^{-fn}}{z_{-}(\mu)} \right] \; ,
\end{equation}
with $z_{\alpha}(x) = \sum_{n} e^{-\varepsilon(n) + \alpha f n  + x n}$ ($\alpha = \pm$) other normalisation constants; and finally, the chemical potential $\mu^{\rm iso}(\rho)$ of the isolated system \cite{guioth2018large,GuiothTBP19} coupled to the overall density $\rho$ through the following equation of state
\begin{equation}\label{eq:density_chempot_relation_MTM}
  \rho = \frac{1}{2} \left[ \frac{z_{+}'\left(\mu^{\rm iso}(\rho)\right)}{z_{+}\left(\mu^{\rm iso}(\rho)\right)} + \frac{z_{-}'\left(\mu^{\rm iso}(\rho)\right)}{z_{-}\left(\mu^{\rm iso}(\rho)\right)} \right] \, .
\end{equation}

In the following, we consider two such models that are allowed to exchange particles through a local contact dynamics (see \cite{GuiothTBP19} for the detailed implementation of the contact).

\subsection{Simultaneous exchange of several particles}

An exchange dynamics where only one particle can be exchanged at a time between the two systems leads to macroscopic detailed balance, which is one of the conditions required for the definition of a non-equilibrium chemical potential \cite{guioth2018large,GuiothTBP19}.
However, one may also consider a more general situation in which more than one particle can be exchanged between the two systems in contact.
In our specific model, if one assumes for instance that $n_{\mathrm{max}}=2$ in each system $A$ and $B$, then at most two particles can be exchanged along one link.

Noting $i_{A}$ and $j_{B}$ the two sites involved in the contact between systems $A$ and $B$, we generally write $T_{c}(n_{i_{A}}', n_{j_{B}}' | n_{i_{A}}, n_{j_{B}})$ the transition rate at contact which satisfies the following local detailed balance condition
\begin{equation}\label{eq:local_db_contact_MTM}
  \ln \frac{ T_{c}(n_{i_{A}}', n_{j_{B}}' | n_{i_{A}}, n_{j_{B}}) }{ T_{c}(n_{i_{A}}, n_{j_{B}} | n_{i_{A}}' , n_{j_{B}}') } = -\left( \Delta \varepsilon_{i_{A}} + \Delta \varepsilon_{j_{B}} \right)
\end{equation}
with $\Delta \varepsilon_{i_{A}} =\varepsilon_{A}(n_{i_{A}}')-\varepsilon_{A}(n_{i_{A}})$ and $\Delta \varepsilon_{j_{B}} =\varepsilon_{B}(n_{j_{B}}') - \varepsilon_{B}(n_{j_{B}})$.

According to the definition of the coarse-grained transition rates in the weak exchange limit \eref{eq:detailed_coarse-grained_transrate_contact}, $\varphi(\rho_{A}, \pm 1)$ read as
\begin{equation}\label{eq:def:varphi_pm1_MTM}
  \fl
  \eqalign{
    \varphi(\rho_{A}, +1) & = T_{c}(1, 0 | 0, 1)P_{A}(0|\rho_{A})P_{B}(1|\rho_{B}) + T_{c}(1, 1 | 0, 2)P_{A}(0|\rho_{A})P_{B}(2|\rho_{B}) \\
    & \quad + T_{c}(2, 0 | 1, 1)P_{A}(1|\rho_{A})P_{B}(1|\rho_{B}) + T_{c}(2, 1 | 1, 2)P_{A}(1|\rho_{A})P_{B}(2|\rho_{B}) \\
    \varphi(\rho_{A}, -1) & = T_{c}(0, 1 | 1, 0)P_{A}(1|\rho_{A})P_{B}(0|\rho_{B}) + T_{c}(0, 2 | 1, 1)P_{A}(1|\rho_{A})P_{B}(1|\rho_{B}) \\
     & \quad + T_{c}(1, 1 | 2, 0)P_{A}(2|\rho_{A})P_{B}(0|\rho_{B}) + T_{c}(1, 2 | 2, 1)P_{A}(2|\rho_{A})P_{B}(1|\rho_{B}) \; ,
   }
\end{equation}
with $P_{k}(n|\rho_{k})$ the single site stationary probability \eref{eq:single_site_probability_distribution_MTM} of system $k$. As for $\varphi(\rho_{A}, \pm 2)$, they read as
\begin{equation}\label{eq:def:varphi_pm2_MTM}
  \eqalign{
    \varphi(\rho_{A}, 2) &= T_{c}(2, 0 | 0, 2) P_{A}(0|\rho_{A}) P_{B}(2|\rho_{B}) \\
    \varphi(\rho_{A}, -2) &= T_{c}(0, 2 | 2, 0) P_{A}(2|\rho_{A}) P_{B}(0|\rho_{B}) \; .
    }    
\end{equation}

\begin{figure}[H]
  \centering
  \includegraphics[width=1.\linewidth]{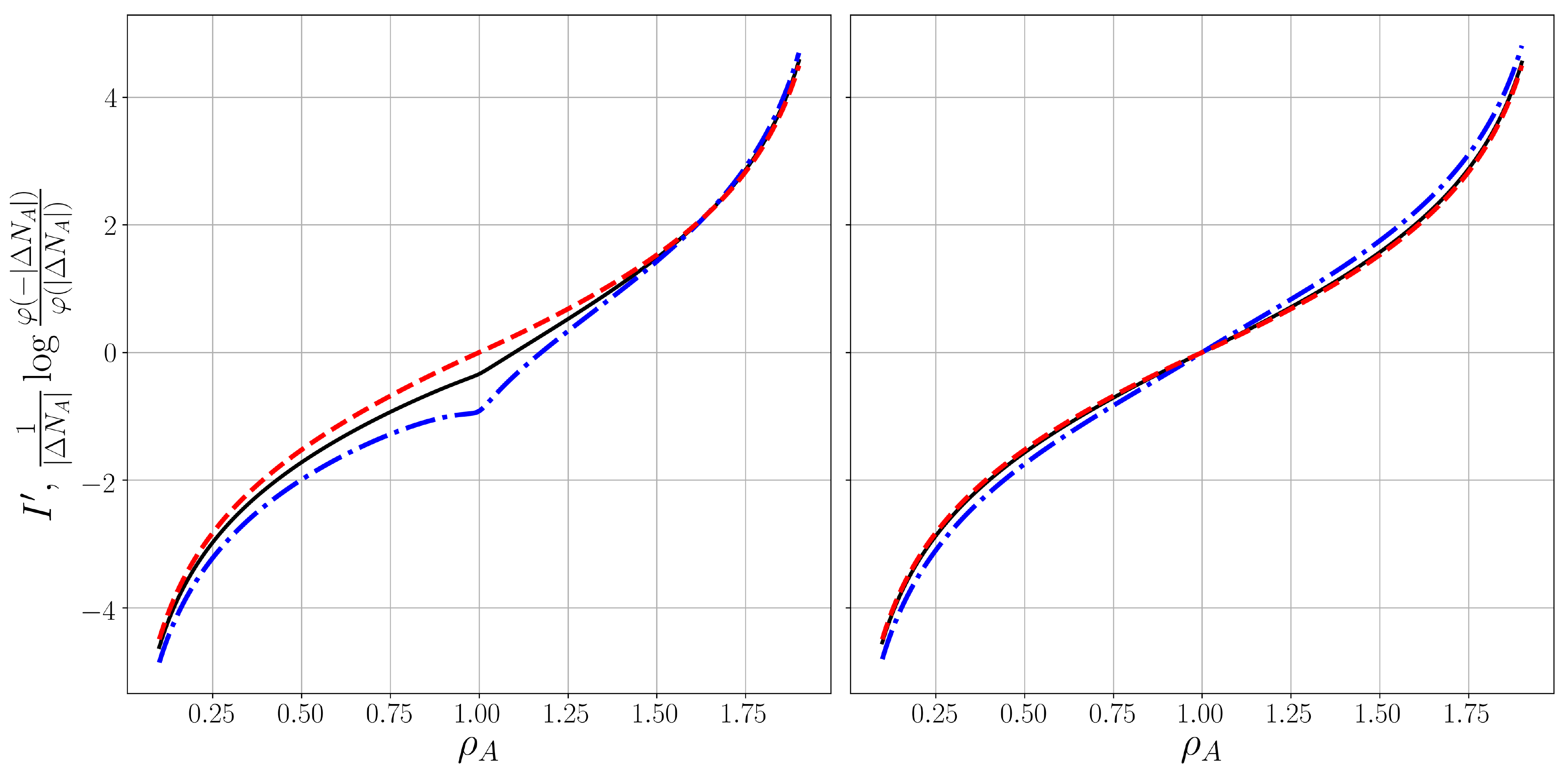}
  \caption{\textbf{Plots of $I'(\rho_{A}|\bar{\rho})$, $\case{1}{|\Delta N_{A}|}\log\left(\varphi(\rho_{A},-|\Delta N_{A}|\right)/\varphi\left(\rho_{A}, |\Delta N_{A}|\right)$ ($|\Delta N_{A}|=1,2$) for ``natural dynamics'' (\emph{left}) and Sasa-Tasaki's rule (\emph{right})}. Black ($\full$) curve: $I'(\rho_{A}|\bar{\rho})$; Blue ($\chain$) curve: $\log(\varphi(\rho_{A},-1)/\varphi(\rho_{A},+1))$; Red ($\broken$) curve: $\case{1}{2}\log(\varphi(\rho_{A},-2)/\varphi(\rho_{A},2))$. The energy functions reads as $\varepsilon_{k}(n_{k})=\varepsilon_{k} n_{k}$ ($k=A,B$). Parameters are: $\varepsilon_{A}=\varepsilon_{B}=1$, $f_{A}=3$, $f_{B}=0$, $\bar{\rho}=1$. One can notice the more significant breaking of the macroscopic detailed balance for the ``natural'' dynamics.}
  \label{fig:Iprime_log_ratio_st_exp}
\end{figure}

Because microscopic detailed balance \eref{eq:DB_violation_micro_contact} is broken, one obtains generally that
\begin{equation}\label{eq:breaking_macro_DB_MTM}
  \ln \frac{\varphi(\rho_{A},-1)}{\varphi(\rho_{A},+1)} \neq \frac{1}{2}\frac{\varphi(\rho_{A},-2)}{\varphi(\rho_{A},+2)} \, .
\end{equation}
We provide numerical estimations of both these ratios for two choices of the dynamics at contact in \fref{fig:Iprime_log_ratio_st_exp}. These are the ``natural dynamics'' that somehow reproduces the bulk transition rates of the isolated systems,
\begin{equation}\label{eq:NatDyn_rule_contact_MTM}
    T_{\mathrm{c}}(n_{i_{A}}',n_{j_{B}}'|n_{i_{A}},n_{j_{B}}) \propto  e^{-\varepsilon_{A}(n_{i_{A}}')} e^{-\varepsilon_{B}(n_{j_{B}}')} 
\end{equation}
and the Sasa-Tasaki's rule that reads
\begin{equation}\label{eq:ST_rule_contact_MTM}
  \fl
    T_{\mathrm{c}}(n_{i_{A}}',n_{j_{B}}'|n_{i_{A}},n_{j_{B}}) \propto
    \cases{
      \exp\left[-\left(\varepsilon_{A}(n_{i_{A}}')-\varepsilon_{A}(n_{i_{A}})\right) \right] & if $n_{i_{A}}' < n_{i_{A}}$ \\
      \exp\left[ -\left(\varepsilon_{B}(n_{j_{B}}') -\varepsilon_{B}(n_{j_{B}}) \right) \right] & if $n_{i_{A}}' > n_{i_{A}}\;~.$
    }
\end{equation}
Exact expressions of the ratios \eref{eq:breaking_macro_DB_MTM} are computed in \ref{sec:appendix_MTM}.

\subsubsection{Computation of the large deviations function $I(\rho_{A}|\bar{\rho})$ in a specific limit}
Since macroscopic detailed balance is broken in the present situation, the large deviations function $I(\rho_{A}|\bar{\rho})$ can only be determined as the solution of the stationary Hamilton-Jacobi equation \eref{eq:hamilton-jacobi_eq}. However, the latter is in general difficult to solve analytically. When two particles can be exchange at most, the Hamilton-Jacobi equation is a polynomial of order $4$ in the variable $e^{I'}$. In order to show that the additivity property is easily broken when macroscopic detailed balance does not hold, we provide here the expression of $I'(\rho_{A}|\bar{\rho})$ when the rate to exchange two particles is much smaller than the one governing the exchange of one particle. To fix the ideas, we rename $\varphi(\rho_{A}, \pm 2) \to \kappa \varphi(\rho_{A}, \pm 2)$ with $\kappa \ll 1$ and $\varphi(\rho_{A},\pm 2) \sim \varphi(\rho_{A}, \pm 1)$.

Noting $I_{\rm MDB}'(\rho_{A}|\bar{\rho})$ the unperturbed solution \jg{that satisfies the macroscopic detailed balance,}
\begin{equation}
  \label{eq:unperturbed_sol_MTM}
  I_{\rm MDB}'(\rho_{A}|\bar{\rho}) = \ln \frac{\varphi(\rho_{A}, -1)}{\varphi(\rho_{A}, 1)}  \; ,
\end{equation}
one obtains, at first order in $\kappa$:
\begin{equation}
  \label{eq:first_order_sol_HJ_MTM_2part}
  \fl
  \eqalign{
    I'(\rho_{A}|\bar{\rho}) & = I_{\rm MDB}'(\rho_{A}|\bar{\rho}) \\
    & \quad + \kappa \frac{\varphi(\rho_{A},2)\left( e^{2I_{\rm MDB}'(\rho_{A}|\bar{\rho})} - 1 \right) + \varphi(\rho_{A},-2)\left(e^{-2I_{\rm MDB}'(\rho_{A}|\bar{\rho})} - 1 \right)}{J_{1}(\rho_{A})}  \\
    & \quad + \Or\left(\kappa^{2}\right) \; ,
    }
\end{equation}
with $J_{1}(\rho_{A}) = \varphi(\rho_{A},1)-\varphi(\rho_{A},-1)$, the macroscopic current associated with the transitions involving an exchange of one particle only. One notes that the expansion stays regular when $\rho_{A}\to \rho_{A}^{\ast \, (1)}$ for which $J_{1}(\rho_{A}^{\ast \, (1)})=0$ and $I_{\rm MDB}'(\rho_{A}^{ \ast \, (1)}) = 0$.

Even if $\varphi(\rho_{A}, \Delta n)$ factorise into two factors as for the ``natural dynamics'' and the Sasa-Tasaki's rule (see \ref{sec:appendix_MTM}), one can easily see with \eref{eq:first_order_sol_HJ_MTM_2part} that a failure (even small) of macroscopic detailed balance is likely to break the additivity property.

\subsection{Continuous mass version of the model}

The above driven lattice gas model was originally introduced in a continuous mass version \cite{guioth2017mass}, that we now briefly discuss in the light of macroscopic detailed balance.
For this continuous mass version of the model, the steady-state distribution takes a form similar to Eq.~(\ref{eq:stationary_probability_MTM}).

To distinguish the continuous mass model from its particle counterpart, we change notations and call $m_{i}\geqslant 0$ the continuous mass at site $i$.
We assume for simplicity that the contact between systems $A$ and $B$ takes place along a single pair of sites $(i_{A}, j_{B})$
For an exchange of a mass $\Delta m $ along this contact link, the coarse-grained transition rate reads, in the weak contact limit, as 
\begin{equation}
  \label{eq:transrate_coarse-grained_weakcontact_limit_continuous_MTM}
  \fl
  \eqalign{
  \varphi(\rho_{A}, \Delta m) & =  \int\limits_{0}^{m_{\mathrm{max}}^{A}} \!\!\!\! \ddr m_{i_{A}} \!\!\! \int\limits_{0}^{m_{\mathrm{max}}^{B}} \! \! \!\!\! \ddr m_{j_{B}} T_{\mathrm{c}}(m_{i_{A}}+ \Delta m, m_{j_{B}}-\Delta m | m_{i_{A}}, m_{j_{B}})   \\
  & \quad \times P(m_{i_{A}}|\rho_{A}) P(m_{j_{B}} | \rho_{B}) \mathbf{1}_{[0,m_{\mathrm{max}}^{A}]}(m_{i_{A}}+\Delta m)  \mathbf{1}_{[0,m_{\mathrm{max}}^{B}]}(m_{j_{B}}- \Delta m)  \; ,
  }
\end{equation}
\jg{where one has introduced $\mathbf{1}_{A}$ the characteristic function of set $A$ such that $\mathbf{1}_{A}(x)=1$ if $x\in A$ and $\mathbf{1}_{A}(x)=0$ otherwise.}
In the following, we focus on the simple case $\varepsilon_{A}(m)=\varepsilon_{A}m$ and $\varepsilon_{B}(m) = \varepsilon_{B} m$.
The single site probability distribution then reads
\begin{equation}
  \label{eq:single_site_proba_distrib_continuous_MTM}
  P(m|\bar{\rho}) = \frac{(\varepsilon - \mu^{\mathrm{iso}}(\bar{\rho}))^{2}-f^{2}}{\varepsilon - \mu^{\mathrm{iso}}(\bar{\rho})} e^{[\mu^{\mathrm{iso}}(\bar{\rho}) - \varepsilon] m } \cosh(fm) 
\end{equation}
%
We focus on the Sasa-Tasaki's contact dynamics (\ref{eq:ST_rule_contact_MTM}), whose generalization to the continuous mass case is straightforward. Assuming that the local mass is unbounded, one finds for the coarse-grained transition rate
\begin{equation}
  \label{eq:transrate_ST_continuous_MTM}
  \fl
      \varphi(\rho_{A}, \Delta m) =
      \cases{
        e^{\mu_{B}^{\mathrm{iso}} \Delta m} \left[ \cosh(f_{B}\Delta m) + \lambda_{B}^{-1}f_{B}\sinh(f_{B} \Delta m) \right] & for $\Delta m > 0$ \\
        e^{\mu_{A}^{\mathrm{iso}} |\Delta m|} \left[ \cosh(f_{A}\Delta m) + \lambda_{A}^{-1}f_{A}\sinh(f_{A} |\Delta m|) \right] & for $\Delta m <0$ 
    }
\nonumber
\end{equation}
with $\lambda_{k}(\rho_{k}) = \varepsilon_{k} - \mu_{k}^{\mathrm{iso}}(\rho_{k})$ ($k=A,B$).

From Eq.~(\ref{eq:transrate_ST_continuous_MTM}), one sees that macroscopic detailed balance does not hold since $\ln [\varphi(\rho_{A},-\Delta m)/\varphi(\rho_{A},\Delta m)]$ is nonlinear in $\Delta m$ for $f_{A}\neq f_{B}$.
Here again, the large deviations function $I(\rho_{A}|\bar{\rho})$ (or, more precisely, its derivative) can thus only be obtained as the solution of the full Hamilton-Jacobi equation (\ref{eq:hamilton-jacobi_eq}), which reads in this case
\begin{eqnarray}
  \label{eq:HJ_eq_continuous_case_MTM}
  &&\int_{\Delta m \geqslant 0} \!\!\! \ddr \Delta m \, \varphi(\rho_{A}, \Delta m )\left( e^{I'(\rho_{A}|\bar{\rho})\Delta m} - 1 \right) \\ \nonumber
  && \qquad \qquad \quad + \int_{\Delta m \geqslant 0} \!\!\! \ddr \Delta m \,  \varphi(\rho_{A},-\Delta m) \, \left(e^{-I'(\rho_{A}|\bar{\rho})\Delta m} -1  \right) = 0  \; .
\end{eqnarray}
Due to the exponential form of the transition rates (\ref{eq:transrate_ST_continuous_MTM}), the two integrals that appear in the Hamilton-Jacobi equation (\ref{eq:HJ_eq_continuous_case_MTM}) can be evaluated explicitly and an algebraic equation over $I'$ can be found.
However, to keep the discussion simple, we have rather performed a perturbative expansion for small values of the driving forces $f_{A}$ and $f_{B}$.
Due to the $f \to -f$ symmetry, the first non-zero order in the expansion is $\Or\left(f^{2}\right)$. At this order, the solution reads
\begin{equation}
  \label{eq:perturbative_sol_HJ_eq_continuous_MTM}
  \eqalign{
   I'(\rho_{A}|\bar{\rho}) &= \mu_{A}^{\mathrm{eq}}(\rho_{A}) - \mu_{B}^{\mathrm{eq}}(\rho_{B}) \\
   & \quad + (f_{A}^{2}-f_{B}^{2})\frac{(\mu_{A}^{\mathrm{eq}})^{2} + \mu_{A}^{\mathrm{eq}}\mu_{B}^{\mathrm{eq}} + (\mu_{B}^{\mathrm{eq}})^{2}}{(\mu_{A}^{\mathrm{eq}})^{2} + 2\mu_{A}^{\mathrm{eq}}\mu_{B}^{\mathrm{eq}} + (\mu_{B}^{\mathrm{eq}})^{2}} + \Or\left(f_{A,B}^{2}\right) \; ,
   }
\end{equation}
with $\mu_{k}^{\rm eq}(\rho_{k}) = \varepsilon_{k} - 1/\rho_{k}$ ($k=A,B$).

One can check explicitly on this example that the large deviations function is not additive as long as $f_{A} \neq f_{B}$.
For $f_{A} = f_{B}$, the large deviations function is additive, at least to order 
$\Or\left(f_{A,B}^{2}\right)$. But in this case, the steady-state densities $\rho_A$ and $\rho_B$ are equal by symmetry, since both systems are identical (except that their size may be different). 

\subsection{In presence of an external potential difference $\Delta U$}

In presence of an external potential difference $\Delta U$ across the contact (see \sref{sec:apply:potential}), only the transition rates at the contact are modified. As shown in a general setting in section \sref{sec:apply:potential}, the general relation between $I'(\rho_{A}|\bar{\rho})$ and $\Delta U$ \eref{eq:ST_relation_DeltaU_large_dev} holds when macroscopic detailed balance is verified.

However, when macroscopic detailed balance is broken, the relation \eref{eq:ST_relation_DeltaU_large_dev} does not hold in general. For instance, in the continuous mass version of the model for which $I'(\rho_{A}|\bar{\rho})$ is given by \eref{eq:perturbative_sol_HJ_eq_continuous_MTM} at first order in the forcing, adding uniform potentials $U_{A}$ and $U_{B}$ on each systems $A$ and $B$ would modify the chemical potentials $\mu_{k}^{\rm eq}$ into $\mu_{k}^{\rm eq} + U_{k}$. As the term proportional to $(f_{A}^{2}-f_{B}^{2})$ in \eref{eq:perturbative_sol_HJ_eq_continuous_MTM} shows, the relation $I' + \Delta U$ \eref{eq:ST_relation_DeltaU_large_dev} no longer holds.

\section{Conclusion}

In this paper, we have focused on the determination and properties of the large deviations function of the density for two steady-state driven systems in contact. This large deviations function turns out to be a convenient tool to determine steady-state densities, that generalizes the notion of chemical potential when the latter cannot be defined, i.e., when the large deviations function is not additive. As shown in \cite{guioth2018large,GuiothTBP19}, this additivity property results both from the macroscopic detailed balance property, and from a factorization property of the contact dynamics.
We have shown how the large deviations function can be evaluated perturbatively when macroscopic detailed balance does not hold, leading generically to a non-additive form of the large deviations function.
In addition, we have provided the large deviations function with a thermodynamic interpretation by generalising the second law of thermodynamics, in the spirit of the Hatano-Sasa relation.
In particular, the measure of the work in the quasi-static limit, as well as the measure of the derivative of the large deviations function through the tilting protocol presented in Sec.~\ref{sec:bias_dyn}, allows one to measure the large deviations function (or its derivative).
Finally, we have illustrated some aspects of this work, like the breaking of macroscopic detailed balance or of the additivity property, on an exactly solvable driven lattice gas model.

Among different open questions, future work may explore the role of a non-vanishing exchange rate at contact, and see how this finite rate may modify the large deviations function of densities. Preliminary results suggest that the additivity property is generically lost when going away from the vanishing exchange rate limit, which makes the characterisation of the coupled systems in terms of large deviations function even more relevant.

\ack
J.G. acknowledges fundings from the Royal Society as well as the French Ministry of Higher Education and Research.


\appendix

\section{Large deviations analysis of density trajectories}
\label{app:path_integral_LD}

We introduce in this appendix ---in a informal way--- the time-reversal symmetry of trajectories of Markov jump processes in a small noise limit. The so-called adjoint dynamics \cite{kipnis2013scaling} is introduced as well.

We first provide an explicit expression of the probability density $\mathcal{P}_{\rho_{A}^{i}}(\omega)$ of a trajectory $\omega = \{\rho_{A}(\tau)\}_{\tau=0}^{t}$ starting at a density $\rho_{A}^{i}$. The latter can be written in the form of a Martin-Siggia-Rose-Janssen-De Dominicis (MSRJD) path integral (see \cite{Janssen1976,PhysRevB.18.4913} in the context of diffusive systems). Path integrals for Poisson processes have often been considered in line with the Doi-Peliti formalism \cite{weber2017master,peliti1985path}. We do not want to enter into these discussions here and we will adopt a formal approach advocated in \cite{andreanov2006field,lefevre2007dynamics,thompson2011lattice} for Markov Jump processes (with the slight difference that the present one is treated at large deviations level):
\begin{equation}
  \label{eq:path_integral_traj}
  \mathcal{P}_{\rho_{A}}(\omega) \underset{V_{A}\to\infty}{\sim} \int \!\! \mathcal{D}\theta \,  e^{- V_{A}\int\limits_{0}^{t} \! \mathrm{d}s \, \left[ \theta(s) \dot{\rho}_{A}(s) - H(\rho_{A}(s),\theta(s)) \right]} \, ,
\end{equation}
where
\begin{equation}
  \label{eq:hamiltonian}
  H(\rho_{A},\theta) = \sum_{\Delta N_{A}} \varphi(\rho_{A},\Delta N_{A})\left( e^{\theta\Delta N_{A}} - 1 \right) \, .
\end{equation} 
The term $S(\omega) = \int_{0}^{t}  \ddr s \, \left[\theta(s) \dot{\rho}_{A}(s) - H(\rho_{A}(s),\theta(s)) \right] $ in the exponential in \eqref{eq:path_integral_traj} is generally referred to as the action of the path $\omega$ in the statistical physics literature \cite{andreanov2006field,lefevre2007dynamics,freidlin1998random,bouchet2016perturbative, cardy1999, peliti1985path}, by analogy with analytical mechanics and its use in Feynman's quantum path integral. Pursuing the mechanical analogy, the function $H$ refers to the Hamiltonian of the stochastic process \cite{andreanov2006field,lefevre2007dynamics}. The presence of an exponential of the response variable $\theta(s)$ in the action is a signature of the underlying Poisson processes \cite{ge2017mathematical}. By contrast, the classical Gaussian noise path integral related to diffusive processes only involves terms up to quadratic order in the response variable $\theta(s)$.

Let us now consider the probability to reach a density $\rho_{A}^{f}$ at large time $t^{f}$, knowing that one starts at $\rho_{A}^{i}$ at time $0$. This conditional probability reads
\begin{equation}
  \label{eq:path_integral_condition_prob}
  P(\rho_{A}^{f},t^{f}|\rho_{A}^{i},0) = \int \!\! \mathcal{D}\rho_{A} \mathcal{P}_{\rho_{A}^{i}}(\omega) \delta\left( \rho_{A}(t^{f}) - \rho_{A}^{f} \right) \; .
\end{equation}
At the thermodynamic limit, only trajectories that minimize the action $S(\omega)$ matter at the leading order in the volume $V$ \cite{tailleur2008mapping,tailleur2007mapping,freidlin1998random,bertini2015macroscopic,bouchet2016perturbative}. They are formally the same as the Hamilton equations and read
\begin{eqnarray}
  \label{eq:instanton_rho}
  \deriv{\rho_{A}}{t} && = \pderiv{H(\rho_{A}(t),\theta(t))}{\theta} \\
                      &&= \sum_{\Delta N_{A}} \Delta N_{A} \varphi(\rho_{A}(t),\Delta N_{A}) e^{\theta(t)\Delta N_{A}}  \nonumber \\
  \deriv{\theta}{t} && = - \pderiv{H(\rho_{A}(t),\theta(t))}{\rho_{A}} \\
               && = \sum_{\Delta N_{A}} \pderiv{\varphi(\rho_{A}(t),\Delta N_{A})}{\rho_{A}} \left(e^{\theta(t)\Delta N_{A}}-1 \right) \nonumber
\end{eqnarray}
with the appropriate boundary conditions, $\rho_{A}(0)=\rho_{A}^{i}$ and \jg{$\rho_{A}(t^{f})=\rho_{A}^{f}$}.

Depending on the boundary conditions, one can distinguish two important situations.

\paragraph{Relaxation dynamics.} The first is at play when the final point $\rho_{A}^{f}=\rho_{A}^{\ast}$ and the initial density $\rho_{A}^{i}$ is a less probable density, one can show that the associated instanton trajectory is simply a relaxation dynamics for which the response variable $\theta(t)$, and thus the action $S$, are uniformly vanishing. According to the Hamilton equations \eqref{eq:instanton_rho}, the relaxation dynamics thus reads, as expected, as
\begin{eqnarray}
  \label{eq:relaxation_dynamics}
  \deriv{\rho_{A}}{t}(t) = J(\rho_{A}(t))&&  = \pderiv{H}{\theta}(\rho_{A}(t),0) \\
                                         && = \sum_{\Delta N_{A}} \Delta N_{A} \varphi(\rho_{A}(t), \Delta N_{A}) \; , \nonumber 
\end{eqnarray}
with $\rho_{A}(0)=\rho_{A}^{i}\neq\rho_{A}^{\ast}$ and $\rho_{A}(t^{f})=\rho_{A}^{\ast}$.

\paragraph{Fluctuation dynamics.} In the opposite case, to realise the time-reversed path that leads to \jg{$\rho_{A}^{f}\neq \rho_{A}^{\ast}$ starting from the stationary state $\rho_{A}(0)=\rho_{A}^{\ast}$}, the system needs to extract energy from the noise and this results in $\theta(s)\neq 0$. Since this trajectory starts at the final point and ends at the initial point of a relaxation dynamics, it should belong to the set of time-reversed trajectories that are associated with relaxation. These trajectories correspond to those of the adjoint dynamics \cite[Appendix 1]{kipnis2013scaling} and it can be shown \cite{bertini2015macroscopic,bouchet2016perturbative} that the most probable trajectory that connects \jg{$\rho_{A}^{\ast}$ to $\rho_{A}^{f}$} matches the most probable trajectory of the adjoint dynamics which corresponds to a dynamics biased by $\theta(t)=I'(\rho_{A}(t)|\bar{\rho})$ at all $t$. 
These rare fluctuating trajectories obey the equation
\begin{eqnarray}
  \label{eq:fluctuation_dynamics}
  \deriv{\rho_{A}}{t}(t) = J^{\dag}(\rho_{A}(t)) && = \pderiv{H}{\theta}(\rho_{A}(t), I'(\rho_{A}(t))) \\
                                                 && = \sum_{\Delta N_{A}} \Delta N_{A} \varphi^{\dag}(\rho_{A}(t), \Delta N_{A}) \; , \nonumber
\end{eqnarray}
with $\rho_{A}(0)=\rho_{A}^{\ast}$, \jg{$\rho_{A}(t^{f})=\rho_{A}^{f}\neq\rho_{A}^{\ast}$} and $\varphi^{\dag}(\rho_{A}, \Delta N_{A}) = \varphi(\rho_{A}, \Delta N_{A})e^{\Delta N_{A} I'(\rho_{A})}$. 

If the macroscopic detailed balance holds, $J=J^{\dag}$, and one recovers that the relaxation path and the fluctuation path are the same, modulo time-reversal. However, when $J\neq J^{\dag}$, \emph{i.e.} when macroscopic detailed balance does not hold, the paths are different. Even if the observation of such fluctuations is certainly difficult in real experiment, this symmetry ---or its absence--- can have important consequences as we have shown in different situations above.
Note that the current can be also expressed in terms of symmetric and anti-symmetric forces, and that the vanishing of $F^{(A)}$ leads to $J=J^{\dag}$ (see \sref{sec:force_and_activity_transrate}).

\section{Work supplied by the potential difference: quasi-static limit and corrections.}
\label{sec:app:quasistatic}

We analyse here what happens when the protocol $\Delta U(s)$ is very slow, \emph{i.e} when $T\to\infty$.
The function $\rho_{A}^{(T)}(s)$ is solution of the time-dependent macroscopic dynamics
\begin{equation}
  \label{eq:time_dependent_macro_dynamics}
  \frac{1}{T} \deriv{\rho_{A}^{(T)}}{s}(s) = J(\Delta U(s) ; \rho_{A}^{(T)}(s)) \, .
\end{equation}
We look for a perturbative solution of the form
\begin{equation}
  \label{eq:exp_solution_time-dependant_macro_dynamics}
  \rho_{A}^{(T)}(s) = \rho_{A}^{(0)}(s) + \frac{1}{T} \rho_{A}^{(1)}(s) + \Or\left(T^{-2}\right) \,.
\end{equation}
Naturally, the quasi-static contribution (zeroth order term of the expansion) reads as
$\rho_{A}^{(0)}(s) = \rho_{A}^{\ast \, \Delta U(s)}$ with, $\rho_{A}^{\ast \, \Delta U(s)}$ solution of $J(\Delta U(s) ; \rho_{A}^{\ast \, \Delta U(s)})=0$. At first order, one finds
\begin{equation}
  \label{eq:perturbative_sol_time-dependant_macro_dynamics}
  \deriv{\rho_{A}^{(0)}(s)}{s} = \rho_{A}^{(1)}(s) \, \pderiv{J}{\rho_{A}}\big(\Delta U(s), \rho_{A}^{(0)}(s)\big)
\end{equation}
which implies
\begin{equation}
\rho_{A}^{(1)}(s) = \deriv{\rho_{A}^{(0)}(s)}{s} \left( \pderiv{J}{\rho_{A}} \right)^{-1}  \; .
\end{equation}
Injecting this expansion in the last term involving $\dot{\mathcal{F}}_{\rm diss}$ in equation (\ref{eq:first_term_average_work_forces_admi_time}) gives, at first order in $T^{-1}$:
\begin{equation}
  \label{eq:first_order_correction_free_energy_dissip_term}
  \fl
 \int_{0}^{1}\!\! \ddr s \, \dot{\mathcal{F}}_{\rm diss}(\Delta U(s), \rho_{A}^{(T)}(s)) = \frac{1}{T} \int_{0}^{1}\,\, \ddr s \, \rho_{A}^{(1)}(s) \pderiv{\dot{\mathcal{F}}_{\rm diss}}{\rho}(\Delta U(s), \rho_{A}^{(0)})  + \Or\left(T^{-2}\right) \,
\end{equation}
where we have used the fact that $\dot{\mathcal{F}}_{\rm diss}(\Delta U(s) ; \rho_{A}^{(0)}(s)) = 0$ since $\rho_{A}^{(0)}(s) = \rho_{A}^{\ast \, \Delta U(s)}$ and that $I_{\Delta U(s)}'$ vanishes at this point.
One eventually gets for the first term of (\ref{eq:first_term_average_work_forces_admi_time})
\begin{equation}\label{eq:quasistatic_limit_first_term_average_work}
  \fl
  \eqalign{
    & \left[ I(\rho_{A}^{(T)}(1)|\bar{\rho}) - I(\rho_{A}^{(T)}(0)|\bar{\rho})\right] + \int_{0}^{1}\!\! \ddr s \, \dot{\mathcal{F}}_{\rm diss}(\Delta U(s), \rho_{A}^{(T)}(s)) \\ 
    & \qquad \qquad \qquad \qquad \qquad \underset{T\to\infty}{\longrightarrow} \lim_{T\to\infty}\left\{ I(\rho_{A}^{(T)}(1)|\bar{\rho}) - I(\rho_{A}^{(T)}(0)|\bar{\rho}) \right\} \; .
}
\end{equation}
We have thus demonstrated that the last term in the right hand side of \eref{eq:first_term_average_work_forces_admi_time} vanishes in the quasi-static limit.

\section{Breaking of macroscopic detailed balance in the exactly solvable model}\label{sec:appendix_MTM}

\subsection{Natural dynamics}
Inserting the exact expressions of the stationary single site distribution \eref{eq:single_site_probability_distribution_MTM} as well as the transition rates at contact \eref{eq:NatDyn_rule_contact_MTM} in \eref{eq:def:varphi_pm1_MTM} and \eref{eq:def:varphi_pm2_MTM}, one obtains
\begin{equation}
  \label{eq:log_phi_pm2_NatDyn_MTM}
  \eqalign{
  \frac{1}{2}\ln\frac{\varphi(\rho_{A}, -2)}{\varphi(\rho_{A}, +2)} &=  \mu_{A}^{\mathrm{iso}}(\rho_{A}) - \mu_{B}^{\mathrm{iso}}(\rho_{B}) \\
  & \quad + \frac{1}{2}\big\{ \upsilon[\mu_{A}^{\mathrm{iso}}, f_{A}](2)-\upsilon[\mu_{B}^{\mathrm{iso}}, f_{B}](2) \big\}  \\
  & \quad - \frac{1}{2} \big\{ \upsilon[\mu_{A}^{\mathrm{iso}}, f_{A}](0)-\upsilon[\mu_{B}^{\mathrm{iso}}, f_{B}](0) \big\} \; . 
  }
\end{equation}
as well as 
\begin{equation}\label{eq:log_phi_pm1_NatDyn_MTM}
  \fl
  \eqalign{
& \ln \frac{\varphi(\rho_{A}, -1)}{\varphi(\rho_{B}, +1)}  = \mu_{A}^{\mathrm{iso}}(\rho_{A}) - \mu_{B}^{\mathrm{iso}}(\rho_{B}) \\
  & \qquad + \ln \left[ \frac{e^{-\varepsilon_{A}(0) - \varepsilon_{A}(1) + \upsilon[\mu_{A}^{\mathrm{iso}}, f_{A}](1)} +  e^{\mu_{A}^{\mathrm{iso}}}e^{-\varepsilon_{A}(1) - \varepsilon_{A}(2) + \upsilon[\mu_{A}^{\mathrm{iso}}, f_{A}](2)}}{e^{-\varepsilon_{A}(0) - \varepsilon_{A}(1) + \upsilon[\mu_{A}^{\mathrm{iso}}, f_{A}](0)} +  e^{\mu_{A}^{\mathrm{iso}}}e^{-\varepsilon_{A}(1) - \varepsilon_{A}(2) + \upsilon[\mu_{A}^{\mathrm{iso}}, f_{A}](1)} } \right]  \\
  & \qquad - \ln \left[ \frac{e^{-\varepsilon_{B}(0) - \varepsilon_{B}(1) + \upsilon[\mu_{B}^{\mathrm{iso}}, f_{B}](1)} +  e^{\mu_{B}^{\mathrm{iso}}}e^{-\varepsilon_{B}(1) - \varepsilon_{B}(2) + \upsilon[\mu_{B}^{\mathrm{iso}}, f_{B}](2)}}{e^{-\varepsilon_{B}(0) - \varepsilon_{B}(1) + \upsilon[\mu_{B}^{\mathrm{iso}}, f_{B}](0)} +  e^{\mu_{B}^{\mathrm{iso}}}e^{-\varepsilon_{B}(1) - \varepsilon_{B}(2) + \upsilon[\mu_{B}^{\mathrm{iso}}, f_{B}](1)}} \right] \; \; .
  }
\end{equation}

\subsection{Sasa-Tasaki's rule}
If now one chooses the Sasa-Tasaki's transition rates \eref{eq:ST_rule_contact_MTM}, one obtains
\begin{equation}
  \label{eq:log_phi_pm2_STdyn_MTM}
  \fl
  \eqalign{
    \ln \frac{\varphi(\rho_{A},-1)}{\varphi(\rho_{A},+1)} &= \mu_{A}^{\rm iso}(\rho_{A}) - \mu_{B}^{\rm iso}(\rho_{B})  \\
    & \quad + \ln \left[ \frac{e^{ -\varepsilon_{A}(0) + \nu\left[ \mu_{A}^{\rm iso}, f_{A} \right](1)  } + e^{\mu_{A}^{\rm iso}}e^{ -\varepsilon_{A}(1) + \nu\left[ \mu_{A}^{\rm iso}, f_{A} \right](2)} }{ e^{ -\varepsilon_{A}(0) + \nu\left[ \mu_{A}^{\rm iso}, f_{A} \right](0)  } + e^{\mu_{A}^{\rm iso}}e^{ -\varepsilon_{A}(1) + \nu\left[ \mu_{A}^{\rm iso}, f_{A} \right](1)} } \right] \\
    & \quad + \ln \left[ \frac{e^{ -\varepsilon_{B}(0) + \nu\left[ \mu_{B}^{\rm iso}, f_{B} \right](1)  } + e^{\mu_{B}^{\rm iso}}e^{ -\varepsilon_{B}(1) + \nu\left[ \mu_{B}^{\rm iso}, f_{B} \right](2)} }{ e^{ -\varepsilon_{B}(0) + \nu\left[ \mu_{B}^{\rm iso}, f_{B} \right](0)  } + e^{\mu_{B}^{\rm iso}}e^{ -\varepsilon_{B}(1) + \nu\left[ \mu_{B}^{\rm iso}, f_{B} \right](1)} } \right] 
    }
\end{equation}
and
\begin{equation}\label{eq:log_phi_pm1_STdyn_MTM}
  \fl
  \eqalign{
    \frac{1}{2}\ln\frac{\varphi(\rho_{A}, -2)}{\varphi(\rho_{A}, +2)} &=  \mu_{A}^{\mathrm{iso}}(\rho_{A}) - \mu_{B}^{\mathrm{iso}}(\rho_{B}) \\
  & \quad + \frac{1}{2}\big\{ \upsilon[\mu_{A}^{\mathrm{iso}}, f_{A}](2)-\upsilon[\mu_{B}^{\mathrm{iso}}, f_{B}](2) \big\}  \\
  & \quad - \frac{1}{2} \big\{ \upsilon[\mu_{A}^{\mathrm{iso}}, f_{A}](0)-\upsilon[\mu_{B}^{\mathrm{iso}}, f_{B}](0) \big\} \; ,
  }
\end{equation}
which appears to be the same as the ``natural'' dynamics \eref{eq:log_phi_pm2_NatDyn_MTM}.


\newpage

\section*{References}

\addcontentsline{toc}{section}{References}


\bibliographystyle{unsrt}
\bibliography{biblio_contact_JSTAT}

\begin{thebibliography}{10}

\bibitem{Oono1998}
Y.~Oono and M.~Paniconi.
\newblock Steady state thermodynamics.
\newblock {\em Prog. Theor. Phys. Supp.}, 130:29, 1998.

\bibitem{sasa2006steady}
S.-i. Sasa and H.~Tasaki.
\newblock Steady state thermodynamics.
\newblock {\em J. Stat. Phys.}, 125(1):125--224, 2006.

\bibitem{Jou03}
J.~Casas-V\'azquez and D.~Jou.
\newblock Temperature in non-equilibrium states: a review of open problems and
  current proposals.
\newblock {\em Rep. Prog. Phys.}, 66:1937, 2003.

\bibitem{Cugliandolo11}
L.~F. Cugliandolo.
\newblock The effective temperature.
\newblock {\em J. Phys. A: Math. Theor.}, 44:483001, 2011.

\bibitem{Levine07}
Y.~Shokef, G.~Shulkind, and D.~Levine.
\newblock Isolated non-equilibrium systems in contact.
\newblock {\em Phys. Rev. E}, 76:030101(R), 2007.

\bibitem{Bertin04}
E.~Bertin, O.~Dauchot, and M.~Droz.
\newblock Temperature in nonequilibrium systems with conserved energy.
\newblock {\em Phys. Rev. Lett.}, 93:230601, 2004.

\bibitem{Martens09}
K.~Martens, E.~Bertin, and M.~Droz.
\newblock Dependence of the fluctuation-dissipation temperature on the choice
  of observable.
\newblock {\em Phys. Rev. Lett.}, 103:260602, 2009.

\bibitem{Dickman2014Inconsistencies}
R.~Dickman and R.~Motai.
\newblock Inconsistencies in steady-state thermodynamics.
\newblock {\em Phys. Rev. E}, 89(3):032134, 2014.

\bibitem{solon2015nat}
A.~P. Solon, Y.~Fily, A.~Baskaran, M.~E. Cates, Y.~Kafri, M.~Kardar, and
  J.~Tailleur.
\newblock Pressure is not a state function for generic active fluids.
\newblock {\em Nature Phys.}, 11(8):673, 2015.

\bibitem{solon2015prl}
A.~P. Solon, J.~Stenhammar, R.~Wittkowski, M.~Kardar, Y.~Kafri, M.~E. Cates,
  and J.~Tailleur.
\newblock Pressure and phase equilibria in interacting brownian spheres.
\newblock {\em Phys. Rev. Lett.}, 114(19):198301, 2015.

\bibitem{winkler2015virial}
R.~G. Winkler, A.~Wysocki, and G.~Gompper.
\newblock Virial pressure in systems of spherical active brownian particles.
\newblock {\em Soft Matter}, 11(33):6680--6691, 2015.

\bibitem{Joyeux2016}
M.~Joyeux and E.~Bertin.
\newblock Pressure of a gas of underdamped active dumbbells.
\newblock {\em Phys. Rev. E}, 93:032605, 2016.

\bibitem{speck2016ideal}
T.~Speck and R.~L. Jack.
\newblock Ideal bulk pressure of active brownian particles.
\newblock {\em Physical Review E}, 93(6):062605, 2016.

\bibitem{fily2018mechanical}
Y.~Fily, Y.~Kafri, A.~P. Solon, J.~Tailleur, and A.~Turner.
\newblock Mechanical pressure and momentum conservation in dry active matter.
\newblock {\em Journal of Physics A: Mathematical and Theoretical},
  51(4):044003, 2018.

\bibitem{bertin2006def}
E.~Bertin, O.~Dauchot, and M.~Droz.
\newblock Definition and relevance of nonequilibrium intensive thermodynamic
  parameters.
\newblock {\em Phys. Rev. Lett.}, 96(12):120601, 2006.

\bibitem{bertin2007intensive}
E.~Bertin, K.~Martens, O.~Dauchot, and M.~Droz.
\newblock Intensive thermodynamic parameters in nonequilibrium systems.
\newblock {\em Phys. Rev. E}, 75(3):031120, 2007.

\bibitem{pradhan2010nonequilibrium}
P.~Pradhan, C.~P. Amann, and U.~Seifert.
\newblock Nonequilibrium steady states in contact: approximate thermodynamic
  structure and zeroth law for driven lattice gases.
\newblock {\em Phys. Rev. Lett.}, 105(15):150601, 2010.

\bibitem{pradhan2011approximate}
P.~Pradhan, R.~Ramsperger, and U.~Seifert.
\newblock Approximate thermodynamic structure for driven lattice gases in
  contact.
\newblock {\em Phys. Rev. E}, 84(4):041104, 2011.

\bibitem{chatterjee2015zeroth}
S.~Chatterjee, P.~Pradhan, and P.K. Mohanty.
\newblock Zeroth law and nonequilibrium thermodynamics for steady states in
  contact.
\newblock {\em Phys. Rev. E}, 91(6):062136, 2015.

\bibitem{guioth2018large}
J.~Guioth and E.~Bertin.
\newblock Large deviations and chemical potential in bulk-driven systems in
  contact.
\newblock {\em Europhys. Lett.}, 123:10002, 2018.

\bibitem{guioth2019lack}
J.~Guioth and E.~Bertin.
\newblock Lack of an equation of state for the nonequilibrium chemical
  potential of gases of active particles in contact.
\newblock {\em J. Chem. Phys.}, 150:094108, 2019.

\bibitem{GuiothTBP19}
J.~Guioth and E.~Bertin.
\newblock Nonequilibrium chemical potentials of steady-state lattice gas models
  in contact: A large-deviation approach.
\newblock {\em Physical Review E}, 100:052125, 2019.

\bibitem{hatano2001steady}
T.~Hatano and S.-i. Sasa.
\newblock Steady-state thermodynamics of langevin systems.
\newblock {\em Phys. Rev. Lett.}, 86:3463, 2001.

\bibitem{bertini2015macroscopic}
L.~Bertini, A.~De~Sole, D.~Gabrielli, G.~Jona-Lasinio, and C.~Landim.
\newblock Macroscopic fluctuation theory.
\newblock {\em Rev. Mod. Phys.}, 87(2):593, 2015.

\bibitem{liggett2012interacting}
T.~M. Liggett.
\newblock {\em Interacting Particle Systems}.
\newblock Grundlehren der mathematischen Wissenschaften. Springer New York,
  2012.

\bibitem{spitzer1970}
F.~Spitzer.
\newblock Interaction of markov processes.
\newblock {\em Adv. Math.}, 5(2):246--290, 1970.

\bibitem{evans2005nonequilibrium}
M.~R. Evans and T.~Hanney.
\newblock Nonequilibrium statistical mechanics of the zero-range process and
  related models.
\newblock {\em J. Phys. A: Math. Gen.}, 38(19):R195, 2005.

\bibitem{Levine2005}
E.~Levine, D.~Mukamel, and G.~M. Sch{\"u}tz.
\newblock Zero-range process with open boundaries.
\newblock {\em J. Stat. Phys.}, 120(5):759--778, Sep 2005.

\bibitem{Evans2006}
M.~R. Evans, S.~N. Majumdar, and R.~K.~P. Zia.
\newblock Canonical analysis of condensation in factorised steady states.
\newblock {\em J. Stat. Phys.}, 123(2):357--390, Apr 2006.

\bibitem{evans2004factorized}
M.~R. Evans, S.~N. Majumdar, and R.~K.~P. Zia.
\newblock Factorized steady states in mass transport models.
\newblock {\em J. Phys. A: Math. Gen.}, 37(25):L275, 2004.

\bibitem{evans2006factorized}
M.~R. Evans, S.~N. Majumdar, and R.~K.~P. Zia.
\newblock Factorized steady states in mass transport models on an arbitrary
  graph.
\newblock {\em J. Phys. A: Math. Gen.}, 39(18):4859, 2006.

\bibitem{zia2004construction}
R.~K.~P. Zia, M.~R. Evans, and S.~N. Majumdar.
\newblock Construction of the factorized steady state distribution in models of
  mass transport.
\newblock {\em J. Stat. Mech.: Theor. Exp.}, 2004(10):L10001, 2004.

\bibitem{derrida2007non}
B.~Derrida.
\newblock Non-equilibrium steady states: fluctuations and large deviations of
  the density and of the current.
\newblock {\em J. Stat. Mech.: Theor. Exp.}, 2007(07):P07023, 2007.

\bibitem{derrida1998asep}
B.~Derrida.
\newblock An exactly soluble non-equilibrium system: The asymmetric simple
  exclusion process.
\newblock {\em Phys. Rep.}, 301(1):65 -- 83, 1998.

\bibitem{katz1984nonequilibrium}
S.~Katz, J.~L. Lebowitz, and H.~Spohn.
\newblock Nonequilibrium steady states of stochastic lattice gas models of fast
  ionic conductors.
\newblock {\em J. Stat. Phys.}, 34(3-4):497--537, 1984.

\bibitem{zia2010twenty}
R.~K.~P. Zia.
\newblock Twenty five years after kls: A celebration of non-equilibrium
  statistical mechanics.
\newblock {\em J. Stat. Phys.}, 138(1-3):20--28, 2010.

\bibitem{maes2003origin}
C.~Maes.
\newblock On the origin and the use of fluctuation relations for the entropy.
\newblock {\em S{\'e}minaire Poincar{\'e}}, 2:29--62, 2003.

\bibitem{maes2003time}
C.~Maes and K.~Neto{\v{c}}n{\`y}.
\newblock Time-reversal and entropy.
\newblock {\em J. Stat. Phys.}, 110(1-2):269--310, 2003.

\bibitem{wynants2010structures}
B.~Wynants.
\newblock {\em Structures of nonequilibrium fluctuations: dissipation and
  activity}.
\newblock PhD thesis, KU Leuven, Belgium, 2010.
\newblock arXiv preprint arXiv:1011.4210.

\bibitem{tasaki2004remark}
H.~Tasaki.
\newblock A remark on the choice of stochastic transition rates in driven
  nonequilibrium systems.
\newblock Preprint arXiv:cond-mat/0407262, 2004.

\bibitem{nayfeh2008perturbation}
A.~H. Nayfeh.
\newblock {\em Perturbation Methods}.
\newblock Physics textbook. Wiley, 2008.

\bibitem{bender1999advanced}
C.M. Bender and S.A. Orszag.
\newblock {\em Advanced Mathematical Methods for Scientists and Engineers I:
  Asymptotic Methods and Perturbation Theory}.
\newblock Springer New York, 1999.

\bibitem{chen1996renormalization}
L.-Y. Chen, N.~Goldenfeld, and Y.~Oono.
\newblock Renormalization group and singular perturbations: Multiple scales,
  boundary layers, and reductive perturbation theory.
\newblock {\em Phys. Rev. E}, 54(1):376, 1996.

\bibitem{chen1994renormalization}
L.~Y. Chen, N.~Goldenfeld, and Y.~Oono.
\newblock Renormalization group theory for global asymptotic analysis.
\newblock {\em Phys. Rev. Lett.}, 73(10):1311, 1994.

\bibitem{oono2012nonlinear}
Y.~Oono.
\newblock {\em The nonlinear world: conceptual analysis and phenomenology}.
\newblock Springer Science \& Business Media, 2012.

\bibitem{touchette2009large}
H.~Touchette.
\newblock The large deviation approach to statistical mechanics.
\newblock {\em Phys. Rep.}, 478(1-3):1--69, 2009.

\bibitem{maes2007static}
C.~Maes and K.~Neto\v{c}n\`{y}.
\newblock Static and dynamical nonequilibrium fluctuations.
\newblock {\em Comptes Rendus Physique}, 8(5):591--597, 2007.

\bibitem{maes2007and}
C.~Maes, K.~Neto{\v{c}}n{\`y}, and B.~Wynants.
\newblock On and beyond entropy production: the case of markov jump processes.
\newblock {\em Markov Proc. Relat. Fields}, 14(3):445--464, 2008.

\bibitem{kaiser2018canonical}
M.~Kaiser, R.~L. Jack, and J.~Zimmer.
\newblock Canonical structure and orthogonality of forces and currents in
  irreversible markov chains.
\newblock {\em J. Stat. Phys.}, 170(6):1019--1050, 2018.

\bibitem{bouchet2016perturbative}
F.~Bouchet, K.~Gawedzki, and C.~Nardini.
\newblock Perturbative calculation of quasi-potential in non-equilibrium
  diffusions: a mean-field example.
\newblock {\em J. Stat. Phys.}, 163(5):1157--1210, 2016.

\bibitem{kipnis2013scaling}
Claude Kipnis and Claudio Landim.
\newblock {\em Scaling limits of interacting particle systems}, volume 320.
\newblock Springer Science \& Business Media, 2013.

\bibitem{bertini2002macroscopic}
L.~Bertini, A.~De~Sole, D.~Gabrielli, G.~Jona-Lasinio, and C.~Landim.
\newblock Macroscopic fluctuation theory for stationary non-equilibrium states.
\newblock {\em J. Stat. Phys.}, 107(3-4):635--675, 2002.

\bibitem{bertini2013clausius}
L.~Bertini, D.~Gabrielli, G.~Jona-Lasinio, and C.~Landim.
\newblock Clausius inequality and optimality of quasistatic transformations for
  nonequilibrium stationary states.
\newblock {\em Phys. Rev. Lett.}, 110(2):020601, 2013.

\bibitem{bertini2012thermodynamic}
L.~Bertini, D.~Gabrielli, G.~Jona-Lasinio, and C.~Landim.
\newblock Thermodynamic transformations of nonequilibrium states.
\newblock {\em J. Stat. Phys.}, 149(5):773--802, 2012.

\bibitem{bertini2015quantitative}
L.~Bertini, A.~De~Sole, D.~Gabrielli, G.~Jona-Lasinio, and C.~Landim.
\newblock Quantitative analysis of the clausius inequality.
\newblock {\em J. Stat. Mech.: Theor. Exp.}, 2015(10):P10018, 2015.

\bibitem{oono1998steady}
Yoshitsugu Oono and Marco Paniconi.
\newblock Steady state thermodynamics.
\newblock {\em Progress of Theoretical Physics Supplement}, 130:29--44, 1998.

\bibitem{ge2017mathematical}
H.~Ge and H.~Qian.
\newblock Mathematical formalism of nonequilibrium thermodynamics for nonlinear
  chemical reaction systems with general rate law.
\newblock {\em J. Stat. Phys.}, 166(1):190--209, 2017.

\bibitem{guioth2017mass}
J.~Guioth and E.~Bertin.
\newblock A mass transport model with a simple non-factorized steady-state
  distribution.
\newblock {\em J. Stat. Mech.: Theor. Exp.}, 2017(6):063201, 2017.

\bibitem{PhysRevLett.95.015702}
E.~Bertin, J.-P. Bouchaud, and F.~Lequeux.
\newblock Subdiffusion and dynamical heterogeneities in a lattice glass model.
\newblock {\em Phys. Rev. Lett.}, 95:015702, 2005.

\bibitem{mclennan1959statistical}
J.~A. McLennan~Jr.
\newblock Statistical mechanics of the steady state.
\newblock {\em Phys. Rev.}, 115(6):1405, 1959.

\bibitem{Janssen1976}
Hans-Karl Janssen.
\newblock On a lagrangean for classical field dynamics and renormalization
  group calculations of dynamical critical properties.
\newblock {\em Zeitschrift f{\"u}r Physik B Condensed Matter}, 23(4):377--380,
  Dec 1976.

\bibitem{PhysRevB.18.4913}
C.~De~Dominicis.
\newblock Dynamics as a substitute for replicas in systems with quenched random
  impurities.
\newblock {\em Physical Review B}, 18:4913--4919, Nov 1978.

\bibitem{weber2017master}
Markus~F. Weber and Erwin Frey.
\newblock Master equations and the theory of stochastic path integrals.
\newblock {\em Reports on Progress in Physics}, 80(4):046601, 2017.

\bibitem{peliti1985path}
L.~Peliti.
\newblock Path integral approach to birth-death processes on a lattice.
\newblock {\em Journal de Physique}, 46(9):1469--1483, 1985.

\bibitem{andreanov2006field}
Alexei Andreanov, Giulio Biroli, Jean-Philippe Bouchaud, and Alexandre Lefevre.
\newblock Field theories and exact stochastic equations for interacting
  particle systems.
\newblock {\em Physical Review E}, 74(3):030101, 2006.

\bibitem{lefevre2007dynamics}
Alexandre Lefevre and Giulio Biroli.
\newblock Dynamics of interacting particle systems: stochastic process and
  field theory.
\newblock {\em Journal of Statistical Mechanics: Theory and Experiment},
  2007(07):P07024, 2007.

\bibitem{thompson2011lattice}
Alasdair~G. Thompson, Julien Tailleur, Michael~E. Cates, and Richard~A. Blythe.
\newblock Lattice models of nonequilibrium bacterial dynamics.
\newblock {\em Journal of Statistical Mechanics: Theory and Experiment},
  2011(02):P02029, 2011.

\bibitem{freidlin1998random}
Mark~Iosifovich Freidlin and Alexander~D. Wentzell.
\newblock {\em Random perturbations of dynamical systems}.
\newblock Springer, 1998.

\bibitem{cardy1999}
John Cardy.
\newblock Field theory and non--equilibrium statistical mechanics.
\newblock Lectures presented as part of the Troisieme Cycle de la Suisse
  Romande, 1999.

\bibitem{tailleur2008mapping}
Julien Tailleur, Jorge Kurchan, and Vivien Lecomte.
\newblock Mapping out-of-equilibrium into equilibrium in one-dimensional
  transport models.
\newblock {\em Journal of Physics A: Mathematical and Theoretical},
  41(50):505001, 2008.

\bibitem{tailleur2007mapping}
Julien Tailleur, Jorge Kurchan, and Vivien Lecomte.
\newblock Mapping nonequilibrium onto equilibrium: the macroscopic fluctuations
  of simple transport models.
\newblock {\em Physical Review Letters}, 99(15):150602, 2007.

\end{thebibliography}

\end{document}